\documentclass[12pt]{article}
\pdfoutput=1
\usepackage{amssymb}
\usepackage{amsmath}
\usepackage{amsthm}
\usepackage{cancel}
\usepackage{slashed}
\usepackage{fullpage}
\usepackage{color}
\usepackage{braket}
\usepackage{graphicx}
\usepackage{multirow}
\usepackage{verbatim}
\usepackage{cite}
\usepackage{bigints}
\usepackage{simplewick}
\usepackage{authblk}

\usepackage{feynmp}
\usepackage{feynmp-auto}
\usepackage{tikz}

\usepackage{caption}
\usepackage{subcaption}
\usepackage{dsfont}
\usepackage{latexsym,amsthm,amssymb,amsmath,amscd,euscript,graphicx,fontenc,eufrak,float, url,hyperref,makecell}
\usepackage{fullpage}

\newcommand{\be}{\begin{equation}}
\newcommand{\ee}{\end{equation}}

\definecolor{darkgreen}{rgb}{0.0,0.5,0.0}
\usepackage{hyperref}
\hypersetup{
    linktocpage,
     colorlinks,
     citecolor=darkgreen,
     linkcolor= darkgreen,
     urlcolor=darkgreen
}

\newcommand{\eps}{\varepsilon}
\newcommand{\cM}{\mathcal M}
\newcommand{\cO}{\mathcal O}
\newcommand{\cMt}{\widetilde \cM}

\newcommand{\st}{\tilde \sigma}
\newcommand{\Z} {Z}
\newcommand{\Gm}{\Gamma_d}
\newcommand{\bbone} { {\mathds 1}}
\title{Infrared Finiteness and Forward Scattering}
\author{Christopher Frye}
\author{Holmfridur Hannesdottir}
\author{\\Nisarga Paul}
\author{Matthew D. Schwartz}
\author{Kai Yan}
\affil{\small \emph{Department of Physics, Harvard University, Cambridge, MA 02138, USA}}
 
 \date{}
\begin{document} 

\begin{fmffile}{feyngraph}
\unitlength = 0.4mm
\maketitle
\thispagestyle{empty}

\begin{abstract}
Infrared divergences have long been heralded to cancel in sufficiently inclusive cross-sections, according to the famous Kinoshita-Lee-Nauenberg theorem
which mandates an initial and final state sum. While well-motivated, this theorem is much weaker than necessary: for finiteness, one need only sum over
initial {\it or} final states. Moreover, the cancellation generically requires the inclusion of the forward scattering process. We provide a number of examples showing the importance of this revised understanding: in $e^+e^- \to Z$ at next-to-leading order, one can sum over certain initial and final states with an arbitrary number of extra photons, or only over final states with a finite number of photons, if forward scattering is included. For Compton scattering, infrared finiteness requires the indistinguishability of hard forward-scattered electrons and photons.  This implies that in addition to experimental limits on the energy and angular resolution, there must also be an experimental limit on the momentum at which electric charge can be observed.  Similar considerations are required to explain why the rate for $\gamma \gamma$ to scatter into photons alone is infrared divergent but the rate for $\gamma \gamma$ to scatter into photons or charged particles is finite.
This new understanding sheds light on the importance of including degenerate initial states in physical predictions, the relevance of disconnected Feynman diagrams, the importance of dressing initial or final-state charged particles, and the quest to properly define the $S$ matrix.
\end{abstract}

\newpage
 \tableofcontents
\newpage
\pagenumbering{arabic}

\section{Introduction}
The appearance and interpretation of infinities has been an essential ingredient of quantum field theory since its inception. 
While ultraviolet divergences appearing in perturbation theory are now completely understood through the program of renormalization, infrared divergences remain somewhat
mysterious. In contrast to ultraviolet divergences, which drop out when amplitudes are expressed directly in terms of other amplitudes, infrared divergences seem only to cancel at the cross-section level for {\it sufficiently inclusive} quantities. It is imperative, therefore, to have a precise definition of sufficiently inclusive, i.e.\ to characterize the minimal set of states that must be included to get a finite cross section.

Part of the reason we find the question of IR finiteness compelling is that its resolution is essential to defining a sensible $S$ matrix. If we define the $S$ matrix in the usual way in quantum field theory, its matrix elements in states of fixed particle number are all infinite at each order in perturbation theory, and zero nonperturbatively. This old problem has not yet limited the applicability of field theory to computing observables at colliders, but is important for studying formal properties of the $S$-matrix, such as its symmetries. The coherent state approach argues that the problem is that isolated charged particles are not well-defined asymptotic states~\cite{Chung:1965zza,Kibble:1969ip,Kulish:1970ut},
but that electrons dressed with a cloud of photons may be. Although the idea is appealing, it is not clear that the dressed/coherent state approach will work for any theory more complicated than QED with massive electrons. If the approach is to succeed it will likely do so through the same mechanism, with the same set of processes, that the cancellation is achieved at the cross section level. Thus, we focus here on cross-section level computations where the the path forward is less obscure.

Although working towards a finite $S$ matrix is a noble goal, there are more practical motivations for understanding IR divergence cancellations. One important one is precision collider physics. Over the last several years, there has been renewed interest in understanding factorization, and its violation in various forms. Consequences of factorization violation include the various large logarithms that appear in perturbative calculations, such as non-global logarithms or super-leading logarithms. Non-global logarithms arise when virtual and real-emission contributions end up in different regions of phase space~\cite{Dasgupta:2001sh,Banfi:2002hw,Kelley:2011tj,Hornig:2011iu,Kelley:2011ng,Schwartz:2014wha,Becher:2016mmh,Larkoski:2015zka}. Super-leading logarithms are associated with collinear factorization violation~\cite{Forshaw:2006fk,Forshaw:2008cq,Keates:2009dn,Schwartz:2017nmr,Schwartz:2018obd}, which is in turn tied to forward scattering, a focus of this paper. Thus, broadly speaking, an improved understanding of IR divergences is relevant to both formal aspects of quantum field theory and  precision collider physics. 

One of the earliest important papers on infrared finiteness was by Bloch and Nordsieck in 1937~\cite{Bloch:1937pw}.  They showed that in QED with massive electrons, infrared singularities in loops and real emission graphs have the same functional form with opposite signs. The Bloch-Nordsieck theorem is that an observable that sums over all possible numbers of final state photons with energies $E< \delta$ is ``sufficiently inclusive," i.e.\ it is infrared finite for any $\delta$. Proofs of the Bloch-Nordsieck theorem were developed sometime later~\cite{Yennie:1961ad,Weinberg:1965nx,Grammer:1973db} and have become textbook material~\cite{Weinberg:1995mt,Muta:2010xua}. 
Essentially, the proof works through the Abelian exponentiation theorem~\cite{Yennie:1961ad,Grammer:1973db}: the soft singularities in QED to all orders in $\alpha$ are given by the exponential of the singularities at 1-loop. With massive electrons, all the singularities in QED are soft in nature and so Abelian exponentiation is all that is needed for the proof.

In theories with massless charged particles, such as QCD, the cancellation of infrared singularities (soft and collinear) is significantly more subtle.
The Bloch-Nordsieck theorem fails in QCD: summing inclusively over final state gluon radiation is insufficient to cancel all infrared singularities, even if the initial state consists only of massive quarks~\cite{Doria:1980ak,DiLieto:1980nkq,Andrasi:1980qw,Carneiro:1980au}. 
Nevertheless, even in QCD infrared divergences 
can be shown to cancel in certain contexts. For example, in hadronic events in $Z$ boson decays, one can identify ``sufficiently inclusive" with ``infrared-and-collinear (IRC) safe": an observable should have the same value if particles with zero energy are added, or if finite energy particles are split into multiple particles going in exactly the same direction. This implies that although the rate for a $Z$ boson to decay to two quarks and nothing else is infinite, the rate for a $Z$ to decay to two ``jets", defined as collections of radiation within an angle $\theta$ including all radiation softer than an energy $\delta$, is well-defined (i.e.\ it is finite)~\cite{Sterman:1977wj}. While this definition of sufficiently inclusive is adequate to remove infrared singularities in $Z \to $ hadron events, it is not a sufficient criterion in other contexts. As we will discuss, $Z\to X$ is special because the $Z\to Z$ forward scattering amplitude is IR finite to all orders. In most other contexts, IRC safety must be generalized.

A non-minimal definition of ``sufficiently inclusive" was proposed by Kinoshita, Lee and Nauenberg in their KLN theorem~\cite{Kinoshita:1962ur,Lee:1964is}. The KLN theorem states that summing over all initial and final states with energies in some compact energy window around a reference energy $E_0$ guarantees finiteness. Stated this way, the theorem is fairly useless as {\it all} states includes $Z$'s, neutrinos, quarks, little red dragons, etc. Fortunately, the KLN theorem derivation involves a sum in a more restricted set: those intermediate particles appearing in any double-sided-cuts~\cite{Kino1950,Kinoshita:1962ur,Naka1958} through any given time-fashioned perturbation theory diagram gives a finite answer. A double sided cut means summing over all possible initial and final states (we give some examples in Section~\ref{sec:klnint}). There are some caveats to the restriction: a number of different diagrams must be included to maintain gauge invariance and Lorentz invariance, but generally the particles involved can be read off the initial graph. 

Despite the importance of the KLN theorem, there are very few explicit computations in the literature showing how the cancellation actually occurs
~\cite{PhysRevD.25.2222,PhysRevD.32.2385,Akhoury:1997pb,Lavelle:2005bt}. One such example was provided by Lavelle and McMullan who showed that IR divergences cancel in processes with an electron scattering off of a background Coulomb potential~\cite{Lavelle:2005bt} (see also~\cite{Lee:1964is,Gastmans:1975sr}). In working out some examples, a number of troublesome features associated with initial state sums emerge. First of all, even though we can define a $n \to m$ cross section mathematically, it is not clear how to think about it physically. Although one can envision a kind of generalization of IRC safely for initial states, including soft and collinear incoming particles, it is not clear how to identify the physical incoming states in a given experiment. In addition, for the KLN cancellation to occur, not only must disconnected diagrams be included, but also an infinite number of photons can participate at any fixed order in the coupling. How to sum this infinite series, with alternating signs for the divergent and finite pieces, requires careful consideration~\cite{Lavelle:2005bt,Khalil:2017yiy}. We examine some of these issues for the process $e^+e^-  + \text{photons} \to Z + \text{photons}$ in Section~\ref{sec:eeZ}. 

Although one can demonstrate the cancellation of IR divergences when summing over initial and final states following the KLN theorem, a careful examination of the proof of theorem provides two revelations: 1) The processes that contribute to assure the cancellation include exactly forward scattering and 2) infrared divergences cancel when summing over final states alone for fixed initial state {\it or} summing over initial states for a fixed final state. This second point is a relief: one can avoid the troublesome aspects of initial state sums. The first point is less of a relief: it requires us to revise our intuition for what states are physically distinguishable. For example, to resolve infrared divergences in $\gamma \gamma \to e^+ e^- $ one must also include $\gamma \gamma$ in the final state. In the end, it seems there are multiple ways to achieve finiteness for this process at next-to-leading order: a final state sum, an initial state sum, or a partial final and initial state sum. We discuss this example in depth in in Section~\ref{sec:eeZ} and some related QED processes are considered in Sections~\ref{sec:compton} and \ref{sec:photon}. A summary of the various results of this paper and some additional thoughts are presented in Section~\ref{sec:conc}.  Appendix~\ref{app:onshell} shows how to compute diagrams with on-shell intermediate states that occur from cuts with disconnected pieces.  Appendix~\ref{app:hemispheres} gives some details of an initial-state jet mass calculation from Section~\ref{sec:eeZ}.

\section{KLN theorem revisited}
We begin by reviewing the KLN theorem and showing that the initial state sum is not necessary. The KLN theorem is attributed to two papers~\cite{Kinoshita:1962ur}, the first by Kinoshita and the second by Lee and Nauenberg~\cite{Lee:1964is}. The Kinoshita paper follows after a paper by Kinoshita and Sirlin~\cite{Kinoshita:1958ru} that considered muon decay $\mu^- \to e^- \nu_\mu \bar{\nu}_e$ in the 4-Fermi theory. They observed that while the exclusive cross section for this process is infrared divergent in the limit of a massless electron, the inclusive cross section is finite when the virtual contribution in $\mu^- \to e^- \nu_\mu \bar{\nu}_e$ is combined with the $\mu^- \to e^- \nu_\mu \bar{\nu}_e \gamma$ cross section. The Kinoshita paper, which builds on work by Nakanishi~\cite{Naka1958}, proves the finiteness of $\mu^- \to e^- \nu_\mu \bar{\nu}_e$  with $m_e=0$ to all orders in perturbative QED. In this way, it generalizes Bloch-Nordsieck to include collinear divergences (mass singularities) as well as soft divergences. Kinoshita also discusses the sum over initial and final states as necessary to cancel mass singularities associated with the muon being massless. Lee and Nauenberg (LN) generalize Kinoshita's result, providing a simple proof for any quantum mechanical system that all infrared divergences (soft and collinear) cancel when initial and all degenerate final states are summed over. Since the LN approach is simple and includes Kinoshita's result, we will focus on it here.

The theorem proved by Lee and Nauenberg  is that the transition amplitude squared is IR finite when summed over initial and final states
\be
\sum_{a \in D(E), b \in D(E)} |\langle b | U(\infty,-\infty) | a \rangle |^2 < \infty \label{LNthm}
\ee
Here $a \in D(E)$ means that the energy of the state $a$ is in the range specified by $D(E)$, e.g. $| E_a - E| < \delta$ for some $\delta >0$.  
For Eq.~\eqref{LNthm} to be true, we must define the sum over states $|b\rangle$ to include the state where $|b\rangle = |a \rangle$, i.e.\ the forward scattering contribution. Including $|a\rangle$ in the sum is critical -- without it the proof does not hold.
%\footnote{It is hard to determine whether Lee and Nauenberg intended for the initial state to be included. In~\cite{Lee:1964is} on p. 1551, in the context of applying their proof to QED they say "For each given state $a$, the subset $D(E_a)$ consists of all other states which differ from $a$ only in the number of infrared photons." Although for their example, potential scattering, including $|b\rangle = |a\rangle$ is  not necessary (as for $e^+e^- \to Z$ in Section~\ref{sec:ifsum} below), for other QED processes, such as Compton scattering (Section~\ref{sec:compton} below) including $|b\rangle = |a\rangle$ is  essential.}

 The operator $U(t_2,t_1)$ is the unitary operator that evolves the system from time $t_1$ to $t_2$. In the interaction picture we write the Hamiltonian as $H(t)=H_0 + V(t)$ with $H_0$ the free Hamiltonian of which the Fock-states $|a\rangle$ are eigenstates and 
\be
U(t_2, t_1 ) = {\mathcal T} \left\{ \exp \left[-i \int_{t_1}^{t_2} dt' V_I(t') \right] \right\} \label{Udef}
\ee
with $V_I(t)=e^{i H_0 (t-t_1)}V(t_1) e^{-i H_0(t-t_1)}$ the interaction picture potential and $t_0$ an arbitrary reference time.

To prove Eq.~\eqref{LNthm}, LN observe that 
\begin{align} \label{bUa}
 |\langle b | U(\infty,-\infty) | a \rangle |^2 &=
  \sum_{i,j}
 \Big[ \langle b | U(\infty,0) | j \rangle \langle  j | U(0,-\infty) | a \rangle \Big] 
 \Big[ \langle b | U(\infty,0) | i \rangle \langle i| U(0,-\infty) | a \rangle\Big]^\star \nonumber \\
 &= R^+_{bij } R_{a ij}^-
\end{align}
with
\begin{align}
R_{bij}^+ &=  \langle i | U(0,\infty) | b \rangle\langle b | U(\infty,0) | j \rangle\\
R_{aij}^-  &=  \langle j | U(0,-\infty) | a \rangle\langle  a | U(-\infty,0) | i \rangle 
\end{align}
So, it is enough to show that 
\be
R^+_{ij} = \sum_{b \in D(E)} R_{bij}^+ < \infty,
\quad
R^-_{ij} = \sum_{a\in D(E)} R_{aij }^- < \infty, \label{Rfinite}
\ee
for Eq. \eqref{LNthm} to hold.

 LN prove Eq.~\eqref{Rfinite} inductively on the number of singular intermediate states. The singularities come from time-ordered-perturbation theory propagators of the form $\frac{1}{E_i - E_j \pm i \eps}$ when $E_i=E_j$. 
By unitarity, we get a finite answer by summing over all states $b$:
\be
R_{ij,\text{all}}^+ =\sum_b  \langle i | U(0,\infty) | b \rangle\langle b | U(\infty,0) | j \rangle = \delta_{ij} < \infty \label{Uall}
\ee
 Thus to show Eq.~\eqref{Rfinite}, we only need to consider states $b$ with energy outside of $D(E)$. But any contribution with $E_b \ne E_i$ or $E_b \ne E_j$ must have at least one non-singular propagator. In this way, LN reduce the number of singular propagators using unitarity and are thereby able to use mathematical induction to complete their proof.  

\section{Final or initial state sums only \label{sec:fsronly}}
The key step in the LN proof is the employment of unitarity, in Eq.~\eqref{Uall}. Note that the sum over all states $b$ includes the intermediate state where $|b\rangle = |i\rangle$, namely forward scattering. Once we accept that forward scattering must be included in the sum, 
we can prove a stronger result than the KLN theorem.
Say we have an initial state $|a\rangle$, at $t=-\infty$ with energy $E$. Then the rate to produce any final states $\langle b|$ in an energy window $D(E)$ around $E$
is finite:
\be
R_{a b}^E =\sum_{j, E_j \in D(E)}   \langle b | U(-\infty,\infty) | j \rangle   \langle j | U(\infty,-\infty) | a \rangle  < \infty \label{RabE}
\ee
To prove this, we only need unitarity and energy conservation. Note that for the LN theorem, the matrix element  $\langle b | U(\infty,0) | a \rangle$ appeared. This matrix element
can be non-vanishing when $E_a=E_b$. Thus the restriction to an energy window was non-trivial. Since  $\langle b | U(\infty, -\infty) | a \rangle \propto \delta(E_b-E_a)$, energy must be conserved and the restriction on $E_j \in D(E)$ is the same as summing over all states. Removing the restriction, we then find 
\be
R_{a b}^E =\sum_{j}   \langle b | U(-\infty,\infty) | j \rangle   \langle j | U(\infty,-\infty) | a \rangle = \delta_{ab}
\ee
which is finite. Note that we are not trying to make this trivial proof seem more complicated than it is -- it really does just require completeness and unitarity. This is in contrast to the LN proof, which is less simple because of the required induction step due to energy non-conservation at finite time.

In words, we have shown that
 \begin{itemize}
 \item For a given fixed initial state $|a\rangle$, the cross section for $|a\rangle$ to go to anything is IR finite.
 \end{itemize}
An analogous proof shows that  
 \begin{itemize}
 \item For a given fixed final state $|b\rangle$, the cross section for anything to go to $|b\rangle$ is IR finite.
 \end{itemize}
Note that in both cases, the states summed over include when the initial state and final state are the same, i.e.\ forward scattering. Importantly, however, we do not need to sum over final {\it and} initial states for IR finiteness. 

Obviously, we do not want to sum over all possible states all the time: the probability for {\it anything}, including both something and nothing, to happen is 1. To get a physical prediction, we must remove a set of states from the sum whose production cross section is finite on its own. The question is then, what is the minimal sufficiently inclusive set of final states required for a finite cross section? 
For perturbative unitarity to hold, the virtual states summed over in the loops must be the same as the real states summed over in the final state phase space integrals. 
The criteria for IR finiteness is therefore the same: any particles with any spins or momenta contributing to the IR divergences in a loop must be included in the phase-space sum.
In the next sections, we will study some particular examples where various subtleties in this requirement emerge.

\section{$\Z \to e^+ e^- (+\gamma)$}
As a warm-up, let us review  the textbook story for the finiteness of $\Z \to e^+e^- (+\gamma)$ in QED with a massless electron. In this and the following sections we always take the electron to be massless, since the massive electron QED case is entirely solved by Bloch-Nordsieck. 

In $d=4-2\eps$ dimensions, the virtual graphs give
\be
\sigma_V =
\hspace{-1mm}
\begin{gathered}
\begin{tikzpicture}
 \node at (-0.05,0) {
\resizebox{20mm}{!}{
     \fmfframe(0,0)(0,0){
\begin{fmfgraph*}(80,80)
\fmfleft{L1}
\fmfright{R1,R2}
\fmf{dashes,tension=2}{L1,v2}
\fmf{fermion}{R2,v1}
\fmf{fermion}{v1,v2}
\fmf{fermion}{v2,v3}
\fmf{fermion}{v3,R1}
\fmffreeze
\fmf{photon}{v1,v3}
\end{fmfgraph*}
}}};
\draw[dashed,red,thick] (0.9,0.8) -- (0.9,-0.8);
\end{tikzpicture}
\end{gathered}
%\hspace{-0.5cm}
%\times
\hspace{-0.5cm}
\begin{gathered}
\begin{tikzpicture}
 \node at (-0.05,0) {
\resizebox{20mm}{!}{
\fmfframe(0,0)(0,0){
\begin{fmfgraph*}(80,80)
\fmfleft{L1,L2}
\fmfright{R1}
\fmf{dashes,tension=2}{R1,v1}
\fmf{fermion}{L1,v1}
\fmf{fermion}{v1,L2}
\end{fmfgraph*}
}}};
\end{tikzpicture}
\end{gathered}
+
\begin{gathered}
\begin{tikzpicture}
 \node at (-0.05,0) {
\resizebox{20mm}{!}{
     \fmfframe(0,0)(0,0){
\begin{fmfgraph*}(80,80)
\fmfleft{L1}
\fmfright{R1,R2}
\fmf{dashes,tension=2}{L1,v1}
\fmf{fermion}{R1,v1}
\fmf{fermion}{v1,R2}
\end{fmfgraph*}
}}};
\draw[dashed,red,thick] (0.9,0.8) -- (0.9,-0.8);
\end{tikzpicture}
\end{gathered}
\hspace{-0.5cm}
\begin{gathered}
\begin{tikzpicture}
 \node at (-0.05,0) {
\resizebox{20mm}{!}{
     \fmfframe(0,0)(0,0){
\begin{fmfgraph*}(80,80)
\fmfleft{L1,L2}
\fmfright{R1}
\fmf{dashes,tension=2}{R1,v2}
\fmf{fermion}{L2,v1}
\fmf{fermion}{v1,v2}
\fmf{fermion}{v2,v3}
\fmf{fermion}{v3,L1}
\fmffreeze
\fmf{photon}{v1,v3}
\end{fmfgraph*}
}}};
\end{tikzpicture}
\end{gathered}
=
\sigma_0^d \Gm \frac{e^2}{\pi^2} \left\{-\frac{1}{4\eps^2} - \frac{3}{8\eps} + \frac{7\pi^2}{48} - 1 \right\}
\delta(1-z)
\label{Zeeloop}
\ee
where $\Gm = \left( \frac { 4 \pi e ^ { - \gamma_E } \mu ^ { 2 } } { Q ^ { 2 } } \right) ^ { \frac { 4 - d } { 2 } }$, 
$\sigma_0^d = \sigma_0 \frac{d-2}{2}\mu^{4 - d}$, and
 $\sigma_0 =\frac{4 \pi g^2}{Q^2}$ with $\sigma_0 \delta(1-z)$ the tree-level
cross section at center-of-mass energy $Q$. We have written the result in terms of $z= \frac{m_Z^2}{Q^2}$ for later convenience. Note that this $1\to2$ process only has support at $z=1$. The real emission graphs give
\be
\sigma_R =
\left|
\begin{gathered}
\begin{tikzpicture}
 \node at (-0.05,0) {
\resizebox{20mm}{!}{
     \fmfframe(0,0)(0,0){
\begin{fmfgraph*}(80,80)
\fmfleft{L1}
\fmfright{R1,R2,R3}
\fmf{dashes,tension=2}{L1,v2}
\fmf{fermion}{R3,v2}
\fmf{phantom}{v2,R1}
\fmffreeze
\fmf{fermion}{v2,v3}
\fmf{fermion}{v3,R1}
\fmffreeze
\fmf{photon}{v3,R2}
\end{fmfgraph*}
}}};
\end{tikzpicture}
\end{gathered}
+
\begin{gathered}
\begin{tikzpicture}
 \node at (-0.05,0) {
\resizebox{20mm}{!}{
     \fmfframe(0,0)(0,0){
\begin{fmfgraph*}(80,80)
\fmfleft{L1}
\fmfright{R1,R2,R3}
\fmf{dashes,tension=2}{L1,v2}
\fmf{phantom}{R3,v2}
\fmf{fermion}{v2,R1}
\fmffreeze
\fmf{fermion}{R3,v3}
\fmf{fermion}{v3,v2}
\fmffreeze
\fmf{photon}{v3,R2}
\end{fmfgraph*}
}}};
\end{tikzpicture}
\end{gathered}
\right|^2
=
\sigma_0^d \Gm \frac{e^2}{\pi^2} \left\{\frac{1}{4\eps^2} + \frac{3}{8\eps}  -  \frac{7\pi^2}{48} + \frac{19}{16}\right\}\delta(1-z)
\label{Zeeg}
\ee
The IR singularities cancel between these two, giving the textbook result $\sigma_V + \sigma_R = \sigma_0 \frac{3e^2}{16\pi^2} \delta(1-z)$. 

Note that for this process,  the cross section is finite without including the $Z\to Z$ forward scattering contribution $\Z \to \Z$. Indeed, the forward-scattering amplitude for $\Z \to \Z$ is IR-finite to all orders in perturbation theory. This follows from the Kinoshita-Poggio-Quinn theorem~\cite{Kinoshita:1962ur,Kinoshita:1975ie,Poggio:1976qr,Sterman:1976jh,Muta:2010xua}. It is also easy to see from general features of infrared divergences~\cite{Landau:1959fi,Libby:1978qf,Sterman:1978bi,Collins:1981ta,Feige:2013zla,Feige:2014wja}: there are no massless external states, so there are no collinear divergences and the external lines are not charged, so there are no soft divergences. 

Note that unitarity alone does not guarantee that these diagrams together are infrared finite. Strictly speaking, unitary holds when summing over cuts of a fixed topology only if the on-shell states in the cut correspond to the those in the loop. In a covariant gauge, the photon propagator does not represent the sum over physical states. Thus only when a gauge invariant combination of all the relevant topologies is summed will unitarity hold. In this case, graphs with external-leg self-energy contributions are amputated while cuts though them (the individual graphs-squared in Eq.~\eqref{Zeeg}) are included. This is the correct procedure as dictated by the LSZ reduction theorem, and the final result is gauge-invariant and subtraction-scheme-independent as it must be.

\section{$e^+e^- \to \Z +X$ \label{sec:eeZ}}
Next let us consider the crossed process, $e^+e^- \to \Z + X$. The virtual graphs are the same as for $ \Z \to e^+e^-$:
\be
\st_{00}=
\hspace{-0.5cm}
\begin{gathered}
\begin{tikzpicture}
 \node at (-0.05,0) {
\resizebox{20mm}{!}{
\fmfframe(0,0)(0,0){
\begin{fmfgraph*}(80,80)
\fmfleft{L1,L2}
\fmfright{R1}
\fmf{dashes,tension=2}{R1,v1}
\fmf{fermion}{L2,v1}
\fmf{fermion}{v1,L1}
\end{fmfgraph*}
}}};
\draw[dashed,red,thick] (1,0.8) -- (1,-0.8);
\end{tikzpicture}
\end{gathered}
\hspace{-2mm}
\begin{gathered}
\begin{tikzpicture}
 \node at (-0.05,0) {
\resizebox{20mm}{!}{
     \fmfframe(0,0)(0,0){
\begin{fmfgraph*}(80,80)
\fmfleft{L1}
\fmfright{R1,R2}
\fmf{dashes,tension=2}{L1,v2}
\fmf{fermion}{R2,v1}
\fmf{fermion}{v1,v2}
\fmf{fermion}{v2,v3}
\fmf{fermion}{v3,R1}
\fmffreeze
\fmf{photon}{v1,v3}
\end{fmfgraph*}
}}};
\end{tikzpicture}
\end{gathered}
\hspace{-3mm}
+
\hspace{-3mm}
\begin{gathered}
\begin{tikzpicture}
 \node at (-0.05,0) {
\resizebox{20mm}{!}{
     \fmfframe(0,0)(0,0){
\begin{fmfgraph*}(80,80)
\fmfleft{L1,L2}
\fmfright{R1}
\fmf{dashes,tension=2}{R1,v2}
\fmf{fermion}{L2,v1}
\fmf{fermion}{v1,v2}
\fmf{fermion}{v2,v3}
\fmf{fermion}{v3,L1}
\fmffreeze
\fmf{photon}{v1,v3}
\end{fmfgraph*}
}}};
\draw[dashed,red,thick] (1.05,0.8) -- (1.05,-0.8);
\end{tikzpicture}
\end{gathered}
%
%\hspace{-0.5cm}
%
\hspace{-2mm}
\begin{gathered}
\begin{tikzpicture}
 \node at (-0.05,0) {
\resizebox{20mm}{!}{
     \fmfframe(0,0)(0,0){
\begin{fmfgraph*}(80,80)
\fmfleft{L1}
\fmfright{R1,R2}
\fmf{dashes,tension=2}{L1,v1}
\fmf{fermion}{R2,v1}
\fmf{fermion}{v1,R1}
\end{fmfgraph*}
}}};
\end{tikzpicture}
\end{gathered}
=
\sigma_0^d \Gm \frac{e^2}{\pi^2} \left\{-\frac{1}{4\eps^2} - \frac{3}{8\epsilon}  + \frac{7\pi^2}{48} -1 \right\}\delta(1-z)
\label{eeZloop}
\ee
where $\Gm = \left( \frac { 4 \pi e ^ { - \gamma_E } \mu ^ { 2 } } { Q ^ { 2 } } \right) ^ { \frac { 4 - d } { 2 } }$, 
$\sigma_0^d = \sigma_0 \frac{d-2}{2}\mu^{4 - d}$, and
 $\sigma_0 =\frac{4 \pi g^2}{Q^2}$ with $\sigma_0 \delta(1-z)$ the tree-level
cross section at center-of-mass energy $Q$. The real emission graphs  give
\begin{multline}
\st_{01}=\left|
\begin{gathered}
\begin{tikzpicture}
 \node at (-0.05,0) {
\resizebox{20mm}{!}{
     \fmfframe(0,0)(0,0){
\begin{fmfgraph*}(80,80)
\fmfleft{L1,L2}
\fmfright{R1,x1,x2,R2,x3,x4,R3}
\fmf{dashes,tension=1.5}{R2,v2}
\fmf{fermion}{L2,v2}
\fmf{phantom}{v2,L1}
\fmffreeze
\fmf{fermion}{v2,v3}
\fmf{fermion}{v3,L1}
\fmffreeze
\fmf{photon}{v3,x1}
\end{fmfgraph*}
}}};
\end{tikzpicture}
\end{gathered}
+
\begin{gathered}
\begin{tikzpicture}
 \node at (-0.05,0) {
\resizebox{20mm}{!}{
 \fmfframe(0,0)(0,0){
\begin{fmfgraph*}(80,80)
\fmfleft{L1,L2}
\fmfright{R1,x1,x2,R2,x3,x4,R3}
\fmf{dashes,tension=1.5}{R2,v2}
\fmf{fermion}{L1,v2}
\fmf{phantom}{L2,v2}
\fmffreeze
\fmf{fermion}{v3,v2}
\fmf{fermion}{L2,v3}
\fmffreeze
\fmf{photon}{v3,x4}
\end{fmfgraph*}
}}};
\end{tikzpicture}
\end{gathered}
\right|^2\\
=\sigma_0^d  \frac{e^2}{\pi^2} \Gm
\left\{
 \delta(1-z) \left(\frac{1}{4\eps^2} - \frac{\pi^2}{16}\right)
 +
  \frac{1+z^2}{4} \left(
  - \frac{1}{\epsilon}
  \left[ \frac{1}{1-z} \right]_+
+
2  \left[ \frac{\ln (1-z)}{1-z} \right]_+ 
%+\left[ \frac{1}{1-z} \right]_+
\right)
  \right\}
  \label{REeeZ}
 \end{multline}
The sum of these graphs does not vanish: $\st_{00} + \st_{01} =\infty$. Here, our notation $\st_{nm}$ refers to the generalized cross section with $n$ incoming photons and $m$ outgoing photons (the generalized cross section is the same as the regular cross section for $2 \to n$ scattering where the incoming particles are massless, see Eq.~\eqref{stdef} below). 

What is different about radiation off incoming and outgoing electrons that changes the singularity structure?
Note that for $\Z \to e^+e^- \gamma$, both the soft and collinear singularities have support only at $z=1$, as can be seen in Eq.~\eqref{Zeeg}. For
$e^+e^- \to \Z \gamma$, if the photon is soft then $z=1$, since a soft photon induces no recoil so the kinematics is the same as for $e^+e^- \to Z$. However, for a hard
collinear photon, additional energy is needed in the final state above that in the $Z$ boson, so we must have $z<1$. Thus the $\frac{1}{\eps} \frac{1}{1-z}$ pole in Eq.~\eqref{REeeZ} is of collinear origin, and different from the $\frac{1}{\eps}\delta(1-z)$ structure of the loop so cannot cancel it.
 That collinear photons are the origin of the difference is consistent with the Bloch-Nordsieck theorem: if the electron were massive, then there would be no collinear singularities and the cross section would be IR finite with either incoming or outgoing electrons. 
 
\subsection{Generalized cross section}
In order to cancel the singularities coming from the loop, we can instead sum over initial states. To sum over initial states, we need a generalization of cross section that can apply to $n\to m$ scattering processes. First of all, we want to allow for forward scattering, so instead of writing $S = 1 + i \cM$, we write
\be
S =  (2\pi)^d \delta^d( P_i^\mu - P_f^\mu)i  \cMt
\ee
so that $\cMt$ includes the forward scattering contribution.\footnote{One might hope that IR finiteness could be achieved using $\cM$ in the conventional way, rather than $\cMt$. Unfortunately, arguments based on cluster decomposition and analyticity that allow us to discard the $\bbone$ in $S$, and more generally the disconnected components, do not  apply with massless particles, when the $S$ matrix is IR divergent. A brief discussion can be found in~\cite[pp. 191-192]{Eden:1966dnq}.
}
 Here $P_i^\mu$ is the sum of all the incoming particles' momenta and $P_f^\mu$ is the sum of all the outgoing particles' momenta. 
Then, rather than computing a cross section, we integrate over both initial and final state phase space. Because the result is Lorentz invariant, it is convenient to work in the center-of-mass frame. So we define
\be
\st \equiv \frac{2^{2d-4} \pi^{2d-2}}{Q^{d-2}} \sum_{\text{spins}} \int d\Pi_i d \Pi_f  | \cMt|^2 (2\pi)^d \delta^d(P_i^\mu - P_f^\mu) \delta^{d-1}(\vec{P}_i + \vec{P}_f) \delta(P_i^0 - Q) \delta^{d-2}(\Omega_{d-1}^{(1)})
\label{stdef}
\ee
where $\Omega_{d-1}^{(1)}$ corresponds to the angle of particle 1 and
\be
d \Pi_i = \prod_{\text{initial states}~j} \frac{d^{d-1} p_j}{(2\pi)^{d-1}} \frac{1}{2 E_{p_j}}, \qquad
d \Pi_f = \prod_{\text{final states}~j} \frac{d^{d-1} p_j}{(2\pi)^{d-1}} \frac{1}{2 E_{p_j}}
\ee
We always sum over initial and final state spins. For a fixed initial or final state, one can always divide by the number of spins to turn the sum into an average. We do not include this averaging factor so that $\st$ corresponds more precisely to what is proven to be infrared finite  in Section~\ref{sec:fsronly}.

The normalization is set so that this definition reduces to the usual definition of a cross section for $2 \to n$ processes where the incoming particles are massless. For example, Eqs.~\eqref{eeZloop} and~\eqref{REeeZ} still hold.

Note that we always sum over spins, for simplicity. One can consider more exclusive cross sections without the spin sum, but since all spins are summed in virtual contributions, we will often need to perform a spin sum to get a finite answer. 

\subsection{Initial and final state sum \label{sec:ifsum}}
Integrating inclusively over the initial state photon phase space at fixed center-of-mass energy $Q$ gives
\be
\st_{10}=\left|
\begin{gathered}
\begin{tikzpicture}
 \node at (-0.05,0) {
\resizebox{20mm}{!}{
     \fmfframe(0,0)(0,0){
\begin{fmfgraph*}(80,80)
\fmfleft{L1,L2,L3}
\fmfright{R1,R2,R3}
\fmf{dashes,tension=1.5}{R2,v2}
\fmf{fermion}{L3,v2}
\fmf{phantom}{v2,L1}
\fmffreeze
\fmf{fermion}{v2,v3}
\fmf{fermion}{v3,L1}
\fmffreeze
\fmf{photon}{v3,L2}
\end{fmfgraph*}
}}};
\end{tikzpicture}
\end{gathered}
+
\begin{gathered}
\begin{tikzpicture}
 \node at (-0.05,0) {
\resizebox{20mm}{!}{
 \fmfframe(0,0)(0,0){
\begin{fmfgraph*}(80,80)
\fmfleft{L1,L2,L3}
\fmfright{R1,R2,R3}
\fmf{dashes,tension=1.5}{R2,v2}
\fmf{fermion}{L1,v2}
\fmf{phantom}{L3,v2}
\fmffreeze
\fmf{fermion}{v3,v2}
\fmf{fermion}{L3,v3}
\fmffreeze
\fmf{photon}{v3,L2}
\end{fmfgraph*}
}}};
\end{tikzpicture}
\end{gathered}
\right|^2
= \sigma_0^d  \frac{e^2}{\pi^2} \Gm \left\{\frac { 1 } { 4\eps^2 } + \frac { 3 } { 8\eps } - \frac { 7 \pi ^ { 2 } } { 48 }  + \frac { 19 } { 16 }\right\}\delta(1-z) 
\ee
This is identical to Eq.~\eqref{Zeeg} and
the infrared divergences (soft and collinear) of these absorption graphs exactly cancel those from the loop in Eq.~\eqref{eeZloop}: $\st_{00} + \st_{10}<\infty$. This is not surprising as we are doing the identical integrals as for $\Z \to e^+ e^- ( +\gamma)$.

Although the  IR divergences of the loop are cancelled by absorption graphs in this way, the emission graphs in Eq.~\eqref{REeeZ} cannot simply be ignored. There is no reason not to include final state radiation in the physical cross section. But since we have already used the loop to cancel the absorption singularities, what is left to cancel them? Since we have now accepted processes with additional photons in the initial state, we should also allow for all such processes. For example, we can have a diagram with
an incoming and outgoing photon interfered with a disconnected graph (the importance of disconnected diagrams has been observed in many contexts~\cite{Lee:1964is,Lavelle:2005bt,Ware:2013zja}). These diagrams give:
\begin{multline}
\st_{11} =
\left(
%
%%graph 1
\begin{gathered}
\begin{tikzpicture}
 \node at (-0.05,0) {
\resizebox{20mm}{!}{
     \fmfframe(0,0)(0,0){
\begin{fmfgraph*}(80,80)
\fmfleft{L1,y1,y2,L2,x1,x2,L3}
\fmfright{R1,w1,w2,R2,z1,z2,R3}
\fmf{dashes,tension=3}{R2,v3}
\fmf{phantom}{L3,v3}
\fmf{phantom}{v3,L1}
\fmffreeze
\fmf{fermion}{L3,v3}
\fmf{fermion}{v3,v2}
\fmf{fermion}{v2,v4}
\fmf{fermion}{v4,L1}
\fmffreeze
\fmf{photon}{v2,w2}
\fmf{photon}{v4,y2}
\end{fmfgraph*}
}}};
\end{tikzpicture}
\end{gathered}
+ \hspace{-3mm}
%%graph 2
\begin{gathered}
\begin{tikzpicture}
 \node at (-0.05,0) {
\resizebox{20mm}{!}{
     \fmfframe(0,0)(0,0){
\begin{fmfgraph*}(80,80)
\fmfleft{L1,y1,y2,L2,x1,x2,L3}
\fmfright{R1,w1,w2,R2,z1,z2,R3}
\fmf{dashes,tension=3}{R2,v2}
\fmf{phantom}{L3,v2}
\fmf{phantom}{v2,L1}
\fmffreeze
\fmf{fermion}{L3,v4}
\fmf{fermion}{v4,v2}
\fmf{fermion}{v2,v3}
\fmf{fermion}{v3,L1}
\fmffreeze
\fmf{photon}{v3,w1}
\fmf{photon}{v4,x1}
\end{fmfgraph*}
}}};
\end{tikzpicture}
\end{gathered}
+ \hspace{-3mm}
%%graph 3
\begin{gathered}
\begin{tikzpicture}
 \node at (-0.05,0) {
\resizebox{20mm}{!}{
     \fmfframe(0,0)(0,0){
\begin{fmfgraph*}(80,80)
\fmfleft{L1,y1,y2,L2,x1,x2,L3}
\fmfright{R1,w1,w2,R2,z1,z2,R3}
\fmf{dashes,tension=3}{R2,v3}
\fmf{phantom}{L3,v3}
\fmf{phantom}{v3,L1}
\fmffreeze
\fmf{fermion}{L3,v3}
\fmf{fermion}{v3,v2}
\fmf{fermion}{v2,v4}
\fmf{fermion}{v4,L1}
\fmffreeze
\fmf{photon}{v2,y2}
\fmf{photon}{v4,w1}
\end{fmfgraph*}
}}};
\end{tikzpicture}
\end{gathered}
\right.
\\
\left.
+ \hspace{-3mm}
%%graph 4
\begin{gathered}
\begin{tikzpicture}
 \node at (-0.05,0) {
\resizebox{20mm}{!}{
     \fmfframe(0,0)(0,0){
\begin{fmfgraph*}(80,80)
\fmfleft{L1,y1,y2,L2,x1,x2,L3}
\fmfright{R1,w1,w2,R2,z1,z2,R3}
\fmf{dashes,tension=3}{R2,v2}
\fmf{phantom}{L3,v2}
\fmf{phantom}{v2,L1}
\fmffreeze
\fmf{fermion}{L3,v4}
\fmf{fermion}{v4,v2}
\fmf{fermion}{v2,v3}
\fmf{fermion}{v3,L1}
\fmffreeze
\fmf{photon}{v3,y2}
\fmf{photon}{v4,z2}
\end{fmfgraph*}
}}};
\end{tikzpicture}
\end{gathered}
+ \hspace{-3mm}
%%graph 5
\begin{gathered}
\begin{tikzpicture}
 \node at (-0.05,0) {
\resizebox{20mm}{!}{
     \fmfframe(0,0)(0,0){
\begin{fmfgraph*}(80,80)
\fmfleft{L1,y1,y2,L2,x1,x2,L3}
\fmfright{R1,w1,w2,R2,z1,z2,R3}
\fmf{dashes,tension=3}{R2,v4}
\fmf{phantom}{L3,v4,L1}
\fmffreeze
\fmf{fermion}{L3,v1,v2,v4,L1}
\fmffreeze
\fmf{photon}{v1,z2}
\fmf{photon}{v2,x1}
\end{fmfgraph*}
}}};
\end{tikzpicture}
\end{gathered}
+ \hspace{-3mm}
%%graph 6
\begin{gathered}
\begin{tikzpicture}
 \node at (-0.05,0) {
\resizebox{20mm}{!}{
     \fmfframe(0,0)(0,0){
\begin{fmfgraph*}(80,80)
\fmfleft{L1,y1,y2,L2,x1,x2,L3}
\fmfright{R1,w1,w2,R2,z1,z2,R3}
\fmf{dashes,tension=3}{R2,v4}
\fmf{phantom}{L3,v4,L1}
\fmffreeze
\fmf{fermion}{L3,v1,v2,v4,L1}
\fmffreeze
\fmf{photon}{v1,x1}
\fmf{photon}{v2,z2}
\end{fmfgraph*}
}}};
\end{tikzpicture}
\end{gathered}
\right)
%\\
%\times
% disconnected graph
\begin{gathered}
\begin{tikzpicture}
\draw[dashed,red,thick] (-1.1,0.8) -- (-1.1,-0.8);
 \node at (0,0) {
\resizebox{20mm}{!}{
\fmfframe(0,0)(0,0){
\begin{fmfgraph*}(80,80)
\fmfstraight
\fmfleft{L1,L2,L3,L4,L5}
\fmfright{R1,R2,R3,R4,R5}
\fmf{dashes,tension=2}{L4,v1}
\fmf{fermion}{v1,R5}
\fmf{fermion}{R3,v1}
\fmffreeze
\fmf{photon}{L2,R2}
\end{fmfgraph*}
}}};
\end{tikzpicture}
\end{gathered}
+ \text{c.c.}\\
= \sigma_0^d  \frac{e^2}{\pi^2} \Gm \left\{
\delta ( 1 - z ) \left( -\frac { 1 } {2 \eps ^ { 2 } } + \frac { \pi ^ { 2 } } { 8 } \right) -\frac{1-z}{2} \hspace{10cm}
%-  \frac{2}{ \epsilon^2 }  \delta ( 1 - z )  + \frac {  3 z ^ { 2 } - 2 z + 1  } { \epsilon } \left[ \frac { 1 } { 1 - z } \right] _ { + } + \text{finite}
\right.
\\
\left.
+
\frac { 3z^2 - 2z + 1 } { 2 } \left(\frac { 1 } { \eps } \left[ \frac { 1 } { 1 - z } \right] _ { + }
- \ln z \left[ \frac { 1 } { 1 - z } \right] _ { + } - 2  \left[ \frac { \ln \left(1-z \right) } { 1 - z } \right] _ { + } \right)
 \right\}
\label{onetoone}
\end{multline}

Evaluating these diagrams requires some care. Consider the first diagram for example. When interfered with the diagram on the right, the outgoing photon momentum is forced to be the same as the incoming photon momentum. This puts one of the intermediate electron propagators on-shell. Normally, on-shell propagators are amputated, but in this case, the on-shell propagator is internal. Such singular propagators were handled by Lee and Nauenberg by including subleading terms in $\epsilon$ using the $i \epsilon$ prescription in time-ordered perturbation theory~\cite{Lee:1964is}. We find however, that their prescription does not work in our case. An alternative method was suggested by Lavelle and McMullan~\cite{Lavelle:2005bt}. A similar situation also occurs when trying to factorize on-shell top production from decay~\cite{Manohar:2014vxa}. Our approach is most similar to that of~\cite{Manohar:2014vxa}.

To deal with the on-shell intermediate state, we must recall that a propagator $\frac{i}{p^2+i \eps}$ is technically a distribution, defined only after integration. Similarly, the $\delta(p^2)$ putting the cut electron on-shell is also a distribution. The product of these distributions must be treated as a distribution, proportional to $\delta'(p_0-\omega_p)$, as we show with an explicit computation of the first diagram above in Appendix~\ref{app:onshell}. The sum of all the diagrams gives the result in Eq.~\eqref{onetoone}. It is intriguing that one cannot interpret these cut diagrams as the product of an amplitude and a conjugate amplitude: the $\delta'$ distribution is only meaningful under the integral of the cut. 

Once we have allowed for disconnected diagrams, nothing prevents contributions with disconnected photons both in $\cMt$ and $\cMt^\star$, such as
\be
\begin{gathered}
\begin{tikzpicture}
 \node at (0,0) {
\resizebox{20mm}{!}{
\fmfframe(0,0)(0,0){
\begin{fmfgraph*}(80,80)
\fmfstraight
\fmfleft{L1,L2,L3,L4,L5,L6}
\fmfright{R1,R2,R3,R4,R5,R6}
\fmf{phantom}{L2,v2}
\fmf{phantom}{v2,L6}
\fmf{dashes,tension=2}{v2,R4}
\fmf{photon}{L1,R1}
\fmffreeze
\fmf{fermion}{L2,v1}
\fmf{fermion}{v1,v2}
\fmf{fermion}{v2,L6}
\fmffreeze
\fmf{photon}{v1,L4}
\end{fmfgraph*}
}}};
\end{tikzpicture}
\end{gathered}
\begin{gathered}
\begin{tikzpicture}
\draw[dashed,red,thick] (-1.1,0.8) -- (-1.1,-0.8);
 \node at (0,0) {
\resizebox{20mm}{!}{
\fmfframe(0,0)(0,0){
\begin{fmfgraph*}(80,80)
\fmfstraight
\fmfright{L1,L2,L3,L4,L5,L6}
\fmfleft{R1,R2,R3,R4,R5,R6}
\fmf{phantom}{L6,v2}
\fmf{phantom}{v2,L2}
\fmf{dashes,tension=2}{v2,R4}
\fmf{photon}{L1,R1}
\fmffreeze
\fmf{fermion}{L6,v2}
\fmf{fermion}{v2,v1}
\fmf{fermion}{v1,L2}
\fmffreeze
\fmf{photon}{v1,L4}
\end{fmfgraph*}
}}};
\end{tikzpicture}
\end{gathered}
+
\begin{gathered}
\begin{tikzpicture}
 \node at (0,0) {
\resizebox{20mm}{!}{
\fmfframe(0,0)(0,0){
\begin{fmfgraph*}(80,80)
\fmfstraight
\fmfleft{L1,L2,L3,L4,L5,L6}
\fmfright{R1,R2,R3,R4,R5,R6}
\fmf{phantom}{L2,v2}
\fmf{phantom}{v2,L6}
\fmf{dashes,tension=2}{v2,R4}
\fmf{phantom}{L1,R1}
\fmffreeze
\fmf{fermion}{L2,v1}
\fmf{fermion}{v1,v2}
\fmf{fermion}{v2,L6}
\fmffreeze
\fmf{phantom}{v1,L4}
\fmffreeze
\fmf{photon,tension=4}{L4,x1}
\fmf{phantom,tension=3}{x1,x2}
\fmf{photon}{x2,R1}
\fmf{photon,right=0.2}{L1,v1}
\end{fmfgraph*}
}}};
\end{tikzpicture}
\end{gathered}
\begin{gathered}
\begin{tikzpicture}
\draw[dashed,red,thick] (-1.1,0.8) -- (-1.1,-0.8);
 \node at (0,0) {
\resizebox{20mm}{!}{
\fmfframe(0,0)(0,0){
\begin{fmfgraph*}(80,80)
\fmfstraight
\fmfright{L1,L2,L3,L4,L5,L6}
\fmfleft{R1,R2,R3,R4,R5,R6}
\fmf{phantom}{L6,v2}
\fmf{phantom}{v2,L2}
\fmf{dashes,tension=2}{v2,R4}
\fmf{photon}{L1,R1}
\fmffreeze
\fmf{fermion}{L6,v2}
\fmf{fermion}{v2,v1}
\fmf{fermion}{v1,L2}
\fmffreeze
\fmf{photon}{v1,L4}
\end{fmfgraph*}
}}};
\end{tikzpicture}
\end{gathered}
+\cdots
\label{condis}
\ee
We have to be careful in evaluating such graphs. If we contract the disconnected photons with each other, as shown in the first graph,
then an extra $\delta^4(0)$ results. This extra infinity is expected by cluster decomposition as the $S$ matrix must factorize into disconnected non-interfering pieces for separated processes~\cite{Eden:1966dnq}. We are not interested in those contractions here, and indeed they are not required by the KLN theorem, as they do not come from double-cut diagrams (see Section~\ref{sec:klnint} below). Thus, when we draw diagrams like this we refer to only the connected interference component, like the second diagram in Eq.~\eqref{condis}, where the disconnected photon in $\cMt$  contracts with the absorbed photon in $\cMt^\star$ or vice-versa.\footnote{The connected interference component means that the uncut full double-cut diagram, wrapped on a cylinder (see Section~\ref{sec:klnint}), is connected. Thus a contribution to $\st$ can be connected even if the contribution $\cMt$ or $\cMt^\star$ is disconnected.}
Note that the connected component must be gauge-invariant on its own as contributions with different numbers of $\delta$ functions cannot cancel.

Focusing on the connected interference terms, in the center of mass frame, the outgoing $Z$ and $\gamma$ have energies $E_Z =\frac{Q^2+m^2}{2Q}$ and $E_\gamma =\frac{Q^2-m^2}{2Q}$ respectively. Since this photon contracts with the absorbed photon, both photons have the same momentum and so the $e^+e^-$ pair has twice the 3-momentum of the $Z$ and energy $E_{ee} \ge 2 E_\gamma$. For energy to be conserved in the $e^+e^-\gamma \to Z$ subdiagram we must then have $E_Z = 3 E_\gamma$, which only has solution for $Q > \sqrt{2} m_Z$ or equivalently $z> \frac{1}{2}$. This kinematic regime is the one we are interested in anyway as the singularity in the original $e^+e^- \to Z$ loop diagram occurred at $z=1$ and so we want to focus on singularities in the $z \approx 1$ regime.
The result is:
\begin{multline}
\st_{21}
=
\left|
\begin{gathered}
\begin{tikzpicture}
 \node at (0,0) {
\resizebox{20mm}{!}{
\fmfframe(0,0)(0,0){
\begin{fmfgraph*}(80,80)
\fmfstraight
\fmfleft{L1,L2,L3,L4,L5,L6}
\fmfright{R1,R2,R3,R4,R5,R6}
\fmf{phantom}{L6,v2}
\fmf{phantom}{v2,L2}
\fmf{dashes,tension=2}{v2,R4}
\fmf{photon}{L1,R1}
\fmffreeze
\fmf{fermion}{L6,v1}
\fmf{fermion}{v1,v2}
\fmf{fermion}{v2,L2}
\fmffreeze
\fmf{photon}{v1,L4}
\end{fmfgraph*}
}}};
\end{tikzpicture}
\end{gathered}
+
\begin{gathered}
\begin{tikzpicture}
 \node at (0,0) {
\resizebox{20mm}{!}{
\fmfframe(0,0)(0,0){
\begin{fmfgraph*}(80,80)
\fmfstraight
\fmfleft{L1,L2,L3,L4,L5,L6}
\fmfright{R1,R2,R3,R4,R5,R6}
\fmf{phantom}{L6,v2}
\fmf{phantom}{v2,L2}
\fmf{dashes,tension=2}{v2,R4}
\fmf{photon}{L1,R1}
\fmffreeze
\fmf{fermion}{L6,v2}
\fmf{fermion}{v2,v1}
\fmf{fermion}{v1,L2}
\fmffreeze
\fmf{photon}{v1,L4}
\end{fmfgraph*}
}}};
\end{tikzpicture}
\end{gathered}
\right|^2_\text{connected}\\
= \sigma_0^d  \frac{e^2}{\pi^2} \Gm \, \Theta(2z-1)
\left\{  \delta ( 1 - z ) \left( \frac{1}{4\eps^2} - \frac{\pi^2}{16} \right) \hspace{10cm}
%- \frac{1}{\eps} \frac {5z^2 - 4z + 1 } { 4 } \left[ \frac { 1 } { 1 - z } \right] _ { + }
\right.\\
\left.
%+ \frac { 5 z ^ { 2 } - 4 z + 1 } { 4 } \left[ \frac { \ln ( 2 z - 1 )}{1 - z}\right]_+ +  \frac { 5 z ^ { 2 } - 4 z + 1 } { 4 }\left[\frac{1}{1-z} \right]_+ 
+ \frac { 5 z ^ { 2 } - 4 z + 1 } { 4 }\left( 
-\frac{1}{\eps} \left[ \frac { 1 } { 1 - z } \right] _+
+ \ln ( 2 z - 1 ) \left[\frac { 1 }{1 - z}\right]_+
+2  \left[ \frac { \ln(1-z)} { 1 - z } \right] _+
%+\left[ \frac { 1 } { 1 - z } \right] _+
\right)
\right\}
\label{twotoone}
\end{multline}
Note the $\theta$-function enforcing the kinematical limit.
In the kinematic regime we are interested in, $z>\frac{1}{2}$, the IR divergences in the sum of Eqs.~\eqref{REeeZ}, \eqref{onetoone} and \eqref{twotoone} exactly cancel: $\st_{01} + \st_{11} + \st_{21} < \infty$. 

This is, however, not the end of the story. Once we agree that connected interference diagrams involving disconnected photons are allowed, we must also allow for such photons to be added to any of the diagrams we have already included. Since disconnected photons do not change the order in the coupling, one can have an arbitrary number of them.\footnote{One way to understand these multi-photon processes from the KLN theorem is that they original originate from diagrams where the photon wraps around the double-cut cylinder more than once~\cite{Lavelle:2005bt}.} We find for a process with $m$ incoming photons and $n$ outgoing photons, for $z>\frac{1}{2}$ and $n>0$ that
 \begin{multline}
 \st_{mn}  = \sigma_0^d  \frac{e^2}{\pi^2} \Gm \left( \delta _ { m - 1 , n } - 2 \delta _ { m , n } + \delta _ { m + 1 , n } \right) 
 \Big\{
  \delta ( 1 - z ) \left( \frac { 1 } { 4\eps^2 } + \frac{\ln n}{\eps} - \frac { \pi ^ { 2 } } { 16 }  + \frac { \ln^2 n } { 2 }  \right)
 + \frac { 1 - z  } { 4 n ^ { 2 } } \delta _ { m n }
 \\
%  - \frac{1}{\eps} \frac{2 n z ( n - m ( 1 - z ) ) + (1-z)^2}{2 n^2} \left[ \frac { 1 } { 1 - z } \right] _ { + } \\
+
 \frac{2 n z [ n - m ( 1 - z ) ] + (1-z)^2}{4 n^2} \left[ \left(-\frac{1}{\eps}+  \ln  \left(\frac{  n - m ( 1 - z ) }{n^3}\right) \right) \left[ \frac { 1 } { 1 - z } \right] _ { + } 
 + 2 \left[  \frac{\ln \left( 1-z \right)}{1-z}  \right]_{+} \right]
 \Big\}
 \end{multline}
 Note that the number of incoming and outgoing photons can differ by at most 1 at this order in the coupling. At higher order, there will be additional terms in $\st_{mn}$ farther from the diagonal. 
 
 It is easy to check from this formula that the IR divergences cancel for any fixed $n$, i.e.\ $\st_{n-1,n} + \st_{n,n} + \st_{n+1,n}$ is finite. Moreover, we find that if we sum over $m$ first, then the sum over $n$ is convergent. Indeed, at large $n$, the asymptotic behavior is
 \be
 \st_{n - 1 , n } + \st_{n , n} + \st_{ n+1 , n} = \sigma_0^d  \frac{e^2}{\pi^2} \Gm \Big\{ - \frac{(1-z)^3}{6 z^2 n^4} + \cO(\frac{1}{n^6}) \Big\} \label{stot}
 \ee
 which is summable. Unfortunately, the series is not absolutely convergent and thus there is an ambiguity on the answer depending on the order in which the terms are summed~\cite{Lavelle:2005bt,Khalil:2017yiy}. The ambiguity can be easily seen by considering reversing the order of the  sum. Holding the number of initial-state photons fixed, we find that the sum over $m$ of
  $\st_{m,m-1} + \st_{m,m} + \st_{m,m+1}$ is finite and scales like $\frac{1}{m^2}$ at large $m$. However, the sum of the $m=0$ terms, $\st_{00} + \st_{01}$, is IR divergent: this is the original problematic sum $e^+e^- \to Z (+\gamma)$. So summing $n$ first, then $m$ we get infinity, while summing $m$ first, then $n$ we get a finite answer.
  
 Even if one could come up with a consistent justification for how to sum the infinite series of $m \to n$ photon contributions, the physical interpretation is still unsettling. Even the $\st_{10}$ contribution, $e^+e^- \gamma \to Z$ is disturbing. The way we have done the calculation involved integrating over the entire kinematically accessible phase space for the incoming photon, including the region of hard, large-angle (non-collinear) photons. It is hard to justify why such photons should be involved in any experimental measurement of $e^+e^+ \to Z$. Instead, we might try to restrict the integral to some infrared-and-collinear safe region. For example, we can consider scattering only incoming ``jets" with invariant mass less than some cutoff $m$. Details using a hemisphere-jet mass definition are provided in Appendix~\ref{app:hemispheres}. We find that as with the total cross section, the jet mass cross section is also IR finite for any fixed number $n$ of outgoing photons. Although the infinite sum retains the same ordering ambiguity as for the full cross section, it is closer to something that could conceivably be measured. Indeed, one can think of the initial-state jet mass calculation as a matching calculation if we set the jet masses equal to the physical electron mass.   While this line of inquiry might ultimately be fruitful, it is not clear that at higher order in perturbation theory, or in more complicated theories like QCD, the IR divergences will still cancel without including forward scattering.  
 
\subsection{KLN interpretation \label{sec:klnint}}
We saw that summing over $e^+ e^- + m\gamma \to Z + n \gamma$ cross sections was infrared finite when summed over $m$ and $n$. This is exactly the kind of cancellation the KLN theorem predicts: including all degenerate initial and final states guarantees finiteness. Although we found the relevant set of graphs by guessing all the relevant physical processes that might contribute at the same order in perturbation theory the KLN theorem actually tells us which subsets of diagrams should cancel: those coming from the double cuts of the same Feynman graph.\footnote{As noted before, although the proof works diagram-by-diagram, it requires unitarity which only holds if the propagator-numerators are the same as the sum over physical on-shell spin states. This is true for gauge theories in physical gauges (like axial gauge) but not true in covariant gauges (like Feynman gauge). One can work in Feynman gauge as long as all of the diagrams required to ensure gauge invariance are included.}

The KLN theorem says that if we take a particular graph and identify the initial and final states, then all possible cuts of that graph should add up to a finite result. For example, the first $\st_{11}$ graph in Eq.~\eqref{onetoone} can be represented as:
\be
\begin{gathered}
\begin{tikzpicture}
 \node at (0,0) {
\resizebox{30mm}{!}{
     \fmfframe(0,0)(0,0){
\begin{fmfgraph*}(120,100)
\fmfleft{L1,L2,L3,L4,L5,L6,L7}
\fmfright{R1,R2,R3,R4,R5,R6,R7}
\fmf{dashes,tension=3}{v1,v5}
\fmf{fermion}{L7,v1}
\fmf{phantom}{v1,L3}
\fmf{fermion,tension=2}{R7,v5}
\fmf{fermion,tension=2}{v5,R3}
\fmffreeze
\fmf{fermion}{v1,v2}
\fmf{fermion,tension=2}{v2,v3}
\fmf{fermion}{v3,L4}
\fmf{photon}{v3,L1}
\fmffreeze
\fmf{photon}{v2,R1}
\end{fmfgraph*}
}}};
\draw[dashed,red,thick] (0.3,1.2) -- (0.3,-1);
\draw[dotted,blue,thick] (-1.4,1.2) -- (-1.4,-1);
\end{tikzpicture}
\end{gathered}
=
\begin{gathered}
\begin{tikzpicture}
 \node at (-0.05,0) {
\resizebox{20mm}{!}{
     \fmfframe(0,0)(0,0){
\begin{fmfgraph*}(80,80)
\fmfleft{L1,y1,y2,L2,x1,x2,L3}
\fmfright{R1,w1,w2,R2,z1,z2,R3}
\fmf{dashes,tension=3}{R2,v3}
\fmf{phantom}{L3,v3}
\fmf{phantom}{v3,L1}
\fmffreeze
\fmf{fermion}{L3,v3}
\fmf{fermion}{v3,v2}
\fmf{fermion}{v2,v4}
\fmf{fermion}{v4,L1}
\fmffreeze
\fmf{photon}{v2,w2}
\fmf{photon}{v4,y2}
\end{fmfgraph*}
}}};
\end{tikzpicture}
\end{gathered}
\begin{gathered}
\begin{tikzpicture}
\draw[dashed,red,thick] (-1.1,0.8) -- (-1.1,-0.8);
 \node at (0,0) {
\resizebox{20mm}{!}{
\fmfframe(0,0)(0,0){
\begin{fmfgraph*}(80,80)
\fmfstraight
\fmfleft{L1,L2,L3,L4,L5}
\fmfright{R1,R2,R3,R4,R5}
\fmf{dashes,tension=2}{L4,v1}
\fmf{fermion}{v1,R5}
\fmf{fermion}{R3,v1}
\fmffreeze
\fmf{photon}{L2,R2}
\end{fmfgraph*}
}}};
\end{tikzpicture}
\end{gathered}
\label{dcut1}
\ee
where the red dashed line is the usual final-state cut and the blue dotted line represents an initial state cut. The diagram should be viewed as on a cylinder, with the right-hand side identified with the left-hand side. Then, for example, the square of the first real emission graph in Eq.~\eqref{REeeZ} can be drawn as
\be
\begin{gathered}
\begin{tikzpicture}
 \node at (0,0) {
\resizebox{30mm}{!}{
     \fmfframe(0,0)(0,0){
\begin{fmfgraph*}(120,100)
\fmfleft{L1,L2,L3,L4,L5,L6,L7}
\fmfright{R1,R2,R3,R4,R5,R6,R7}
\fmf{dashes,tension=3}{v1,v5}
\fmf{fermion}{L7,v1}
\fmf{phantom}{v1,L3}
\fmf{fermion,tension=2}{R7,v5}
\fmf{fermion,tension=2}{v5,R3}
\fmffreeze
\fmf{fermion}{v1,v2}
\fmf{fermion,tension=2}{v2,v3}
\fmf{fermion}{v3,L4}
\fmf{photon}{v3,L1}
\fmffreeze
\fmf{photon}{v2,R1}
\end{fmfgraph*}
}}};
\draw[dashed,red,thick] (0.3,1.2) -- (0.3,-1);
\draw[dotted,blue,thick] (-0.9,1.2) -- (-0.9,-1);
\end{tikzpicture}
\end{gathered}
=
\begin{gathered}
\begin{tikzpicture}
 \node at (-0.05,0) {
\resizebox{20mm}{!}{
     \fmfframe(0,0)(0,0){
\begin{fmfgraph*}(80,80)
\fmfleft{L1,L2}
\fmfright{R1,R2}
\fmf{fermion}{L2,v1,v2,L1}
\fmf{dashes}{v1,R2}
\fmf{photon}{v2,R1}
\end{fmfgraph*}
}}};
\end{tikzpicture}
\end{gathered}
\begin{gathered}
\begin{tikzpicture}
 \node at (-0.05,0) {
\resizebox{20mm}{!}{
\fmfframe(0,0)(0,0){
\begin{fmfgraph*}(80,80)
\fmfleft{L1,L2}
\fmfright{R1,R2}
\fmf{fermion}{R2,v1,v2,R1}
\fmf{dashes}{v1,L2}
\fmf{photon}{v2,L1}
\end{fmfgraph*}
}}};
\draw[dashed,red,thick] (-1.2,1.2) -- (-1.2,-1);
\end{tikzpicture}
\end{gathered}
\label{dcut2}
\ee
A different double-cut diagram can produce the third diagram in Eq.~\eqref{onetoone} or the disconnected diagram in Eq.~\eqref{twotoone}:
\be
\begin{gathered}
\begin{tikzpicture}
 \node at (-0.05,0) {
\resizebox{20mm}{!}{
     \fmfframe(0,0)(0,0){
\begin{fmfgraph*}(80,80)
\fmfleft{L1,y1,y2,L2,x1,x2,L3}
\fmfright{R1,w1,w2,R2,z1,z2,R3}
\fmf{dashes,tension=3}{v5,v3}
\fmf{phantom}{R3,v5}
\fmf{phantom}{v5,R1}
\fmf{phantom}{L3,v3}
\fmf{phantom}{v3,L1}
\fmffreeze
\fmf{fermion}{z1,v5}
\fmf{fermion}{v5,w2}
\fmffreeze
\fmf{fermion}{L3,v3}
\fmf{fermion}{v3,v2}
\fmf{fermion}{v2,v4}
\fmf{fermion}{v4,L1}
\fmffreeze
\fmf{photon}{v2,y2}
\fmf{photon}{v4,w1}
\end{fmfgraph*}
}}};
\draw[dashed,red,thick] (-0.1,1.2) -- (-0.1,-1);
\draw[dotted,blue,thick] (-1.2,1.2) -- (-1.2,-1);
\end{tikzpicture}
\end{gathered}
=
\begin{gathered}
\begin{tikzpicture}
 \node at (-0.05,0) {
\resizebox{20mm}{!}{
     \fmfframe(0,0)(0,0){
\begin{fmfgraph*}(80,80)
\fmfleft{L1,y1,y2,L2,x1,x2,L3}
\fmfright{R1,w1,w2,R2,z1,z2,R3}
\fmf{dashes,tension=3}{R2,v3}
\fmf{phantom}{L3,v3}
\fmf{phantom}{v3,L1}
\fmffreeze
\fmf{fermion}{L3,v3}
\fmf{fermion}{v3,v2}
\fmf{fermion}{v2,v4}
\fmf{fermion}{v4,L1}
\fmffreeze
\fmf{photon}{v2,y2}
\fmf{photon}{v4,w1}
\end{fmfgraph*}
}}};
\end{tikzpicture}
\end{gathered}
\begin{gathered}
\begin{tikzpicture}
\draw[dashed,red,thick] (-1.1,0.8) -- (-1.1,-0.8);
 \node at (0,0) {
\resizebox{20mm}{!}{
\fmfframe(0,0)(0,0){
\begin{fmfgraph*}(80,80)
\fmfstraight
\fmfleft{L1,L2,L3,L4,L5}
\fmfright{R1,R2,R3,R4,R5}
\fmf{dashes,tension=2}{L4,v1}
\fmf{fermion}{v1,R5}
\fmf{fermion}{R3,v1}
\fmffreeze
\fmf{photon}{L2,R2}
\end{fmfgraph*}
}}};
\end{tikzpicture}
\end{gathered}
%%%%
,
\hspace{1cm}
%%%%%
\begin{gathered}
\begin{tikzpicture}
 \node at (-0.05,0) {
\resizebox{20mm}{!}{
     \fmfframe(0,0)(0,0){
\begin{fmfgraph*}(80,80)
\fmfleft{L1,y1,y2,L2,x1,x2,L3}
\fmfright{R1,w1,w2,R2,z1,z2,R3}
\fmf{dashes,tension=3}{v5,v3}
\fmf{phantom}{R3,v5}
\fmf{phantom}{v5,R1}
\fmf{phantom}{L3,v3}
\fmf{phantom}{v3,L1}
\fmffreeze
\fmf{fermion}{z1,v5}
\fmf{fermion}{v5,w2}
\fmffreeze
\fmf{fermion}{L3,v3}
\fmf{fermion}{v3,v2}
\fmf{fermion}{v2,v4}
\fmf{fermion}{v4,L1}
\fmffreeze
\fmf{photon}{v2,y2}
\fmf{photon}{v4,w1}
\end{fmfgraph*}
}}};
\draw[dashed,red,thick] (-0.1,1.2) -- (-0.1,-1);
\draw[dotted,blue,thick] (-0.62,1.2) -- (-0.62,-1);
\end{tikzpicture}
\end{gathered}
=
\left|
\begin{gathered}
\begin{tikzpicture}
 \node at (0,0) {
\resizebox{20mm}{!}{
\fmfframe(0,0)(0,0){
\begin{fmfgraph*}(80,80)
\fmfstraight
\fmfleft{L1,L2,L3,L4,L5,L6}
\fmfright{R1,R2,R3,R4,R5,R6}
\fmf{phantom}{L6,v2}
\fmf{phantom}{v2,L2}
\fmf{dashes,tension=2}{v2,R4}
\fmf{photon}{L1,R1}
\fmffreeze
\fmf{fermion}{L6,v1}
\fmf{fermion}{v1,v2}
\fmf{fermion}{v2,L2}
\fmffreeze
\fmf{photon}{v1,L4}
\end{fmfgraph*}
}}};
\end{tikzpicture}
\end{gathered}
\right|^2
\ee

Now, as we have observed, the double-cut sum in the KLN theorem also includes contributions where both cuts are in the same place, giving forward scattering contributions. 
For example, the double cut diagram in Eqs.~\eqref{dcut1} and \eqref{dcut2} also generates forward-scattering cuts,
\be
\begin{gathered}
\begin{tikzpicture}
 \node at (0,0) {
\resizebox{30mm}{!}{
     \fmfframe(0,0)(0,0){
\begin{fmfgraph*}(120,100)
\fmfleft{L1,L2,L3,L4,L5,L6,L7}
\fmfright{R1,R2,R3,R4,R5,R6,R7}
\fmf{dashes,tension=3}{v1,v5}
\fmf{fermion}{L7,v1}
\fmf{phantom}{v1,L3}
\fmf{fermion,tension=2}{R7,v5}
\fmf{fermion,tension=2}{v5,R3}
\fmffreeze
\fmf{fermion}{v1,v2}
\fmf{fermion,tension=2}{v2,v3}
\fmf{fermion}{v3,L4}
\fmf{photon}{v3,L1}
\fmffreeze
\fmf{photon}{v2,R1}
\end{fmfgraph*}
}}};
\draw[dashed,red,thick] (0.3,1.2) -- (0.3,-1);
\draw[dotted,blue,thick] (0.2,1.2) -- (0.2,-1);
\end{tikzpicture}
\end{gathered}
=
\begin{gathered}
\begin{tikzpicture}
 \node at (0,0) {
\resizebox{30mm}{!}{
     \fmfframe(0,0)(0,0){
\begin{fmfgraph*}(120,100)
 \fmfleft{L1,L2}
         \fmfright{R1,R2}
        \fmf{phantom}{v1,L1}
        \fmf{fermion}{v1,v2}
        \fmf{dashes}{R2,v4}
        \fmf{photon}{v2,R1}
        \fmf{fermion}{v4,v3}
        \fmf{phantom}{L2,v3}
        \fmf{fermion}{v3,v1}
        \fmf{fermion}{v4,v2}
        \fmffreeze
        \fmf{photon}{v1,va}
        \fmf{phantom,tension=2}{va,vb}
        \fmf{photon}{vb,L1}
        \fmf{dashes}{vd,v3}
        \fmf{phantom,tension=2}{vc,vd}
        \fmf{dashes}{L2,vc}       
\end{fmfgraph*}
}}};
\draw[dashed,red,thick] (-0.95,1) -- (-0.95,-1);
\end{tikzpicture}
\end{gathered}
\ee
%\end{multline}
The double-cut sum also includes contributions where the initial and final state cuts are swapped from Eq.~\eqref{dcut2}
\be
%\begin{multline}
\sigma_{\gamma Z ee}=
\begin{gathered}
\begin{tikzpicture}
 \node at (0,0) {
\resizebox{30mm}{!}{
     \fmfframe(0,0)(0,0){
\begin{fmfgraph*}(120,100)
\fmfleft{L1,L2,L3,L4,L5,L6,L7}
\fmfright{R1,R2,R3,R4,R5,R6,R7}
\fmf{dashes,tension=3}{v1,v5}
\fmf{fermion}{L7,v1}
\fmf{phantom}{v1,L3}
\fmf{fermion,tension=2}{R7,v5}
\fmf{fermion,tension=2}{v5,R3}
\fmffreeze
\fmf{fermion}{v1,v2}
\fmf{fermion,tension=2}{v2,v3}
\fmf{fermion}{v3,L4}
\fmf{photon}{v3,L1}
\fmffreeze
\fmf{photon}{v2,R1}
\end{fmfgraph*}
}}};
\draw[dashed,red,thick] (-0.9,1.2) -- (-0.9,-1);
\draw[dotted,blue,thick]  (0.3,1.2) -- (0.3,-1);
\end{tikzpicture}
\end{gathered}
=
\begin{gathered}
\begin{tikzpicture}
 \node at (-0.05,0) {
\resizebox{20mm}{!}{
     \fmfframe(0,0)(0,0){
\begin{fmfgraph*}(80,80)
\fmfleft{R1,R2}
\fmfright{L1,L2}
\fmf{fermion}{L2,v1,v2,L1}
\fmf{dashes}{v1,R2}
\fmf{photon}{v2,R1}
\end{fmfgraph*}
}}};
\end{tikzpicture}
\end{gathered}
\begin{gathered}
\begin{tikzpicture}
 \node at (-0.05,0) {
\resizebox{20mm}{!}{
\fmfframe(0,0)(0,0){
\begin{fmfgraph*}(80,80)
\fmfleft{R1,R2}
\fmfright{L1,L2}
\fmf{fermion}{R2,v1,v2,R1}
\fmf{dashes}{v1,L2}
\fmf{photon}{v2,L1}
\end{fmfgraph*}
}}};
\draw[dashed,red,thick] (-1.2,1.2) -- (-1.2,-1);
\end{tikzpicture}
\end{gathered}
\ee
The process corresponding to these cuts are $\gamma Z \to \gamma Z$ and  $\gamma Z \to e^+ e^-$ respectively, neither of which seems very much like the original 
$e^+e^- \to Z$ process whose singularities we were trying to cancel. 
%\end{multline}

If the  KLN theorem requires us to sum over this large number of contributions, why do a subset of them cancel among themselves? In some cases, we can find a clear answer.
 For example, we found that the virtual contribution to $e^+e^- \to \Z$ was IR finite when summed with $e^+ e^- \gamma \to \Z$; these are contributions with a fixed final state, namely the $\Z$. 
 Since $\Z \to \Z$  is IR finite on its own, the sum of contributions of anything $\to Z$ will be finite whether or not we include forward scatting.
 
Similarly, we found
 $e^+e^- \to \Z \gamma$ canceled against $\gamma e^+ e^- \to \Z \gamma$ and $\gamma \gamma e^+ e^- \to \Z \gamma$. These contributions all have final states with $\Z \gamma$. Thus we could explain the cancellation among these terms alone if the  forward scattering contribution $\Z  \gamma \to \Z \gamma$ were infrared finite. 
 Evaluating the loop, we find 
   \begin{multline}
 \st_{\gamma Z}=
 \begin{gathered}
\begin{tikzpicture}
 \node at (0,0) {
\resizebox{30mm}{!}{
     \fmfframe(0,0)(0,0){
\begin{fmfgraph*}(120,100)
 \fmfleft{L1,L2}
         \fmfright{R1,R2}
        \fmf{phantom}{v1,L1}
        \fmf{fermion}{v1,v2}
        \fmf{dashes}{R2,v4}
        \fmf{photon}{v2,R1}
        \fmf{fermion}{v4,v3}
        \fmf{phantom}{L2,v3}
        \fmf{fermion}{v3,v1}
        \fmf{fermion}{v2,v4}
        \fmffreeze
        \fmf{photon}{v1,va}
        \fmf{phantom,tension=2}{va,vb}
        \fmf{photon}{vb,L1}
        \fmf{dashes}{vd,v3}
        \fmf{phantom,tension=2}{vc,vd}
        \fmf{dashes}{L2,vc}       
\end{fmfgraph*}
}}};
\draw[dashed,red,thick] (-0.95,1) -- (-0.95,-1);
\end{tikzpicture}
\end{gathered}
+ \text{crossings} + \text{c.c.}\\
=
\sigma_0^d  \frac{e^2}{\pi^2} \Gm
\times \begin{cases}
 \dfrac{5 z ^ 2 - 4 z + 1}{4\left( 1-z \right)} \left[-\dfrac{1}{\eps}+ 2 \ln(1-z)\right]
+ \dfrac{3 z^2-2z+1}{2\left(1-z\right)} \ln z + \dfrac{1-z}{2}
& z<\frac{1}{2}\\[4mm]
\dfrac{ 3 z^2 - 2 z + 1}{2}\ln z \left[ \dfrac{1}{1-z}\right]_++\dfrac{ -5 z^2 + 4 z - 1 }{4}  \ln ( 2 z - 1 ) \left[ \dfrac{1}{1-z} \right]_+ + \dfrac{ 1 - z}{2},
  &z  > \frac{1}{2}
\end{cases}
%\\
%=
%\sigma_0  \frac{e^2}{\pi^2} \Gm\ \frac{ ( 1 -  z )^{ d - 5 } }{ 4 }
%\times \begin{cases}
%- \frac { 1 } { \eps } \left(5 z ^ 2 - 4 z + 1\right)
%+
%\left( 6 z^2 - 4 z + 2 \right) \ln z + 2 ( 1 - z ) ^ { 2 },
%& z<\frac{1}{2}\\[2mm]
% \left( 6 z^2 - 4 z + 2 \right) \ln z + 2 ( 1 - z ) ^ { 2 } - \left( 5 z^2 - 4 z + 1 \right)  \ln ( 2 z - 1 ) - \eps \frac{\pi^2}{3},
%  &z  > \frac{1}{2}
%\end{cases}
\label{gZgZ}
 \end{multline}
In this expression, the $\cO(\eps)$ piece is evaluated at $z=1$ since it only contributes when multiplying the $\frac{1}{\eps} \delta(1-z)$ term from the expansion of the prefactor.

 We see that  the forward scattering contribution is infrared divergent, but only for $z< \frac{1}{2}$ ($Q^2 > 2 m_Z^2$). For $z \approx 1$ ($Q \approx m_Z$), which is the limit in which we can examine the IR divergence associated with $e^+e^- \to Z$, the $\gamma Z \to \gamma Z$ forward scattering contribution is IR finite. This explains why the the sum over all other $X \to \gamma Z$ diagrams, $\sum \st_{n 1}$ will be IR finite, as we have seen. 

At high energy, $Q^2 > 2 m_Z^2$, the forward scattering process is IR divergent. Note however that this is the identical to threshold above
which the $2\to 1$ diagrams $\st_{21}$ vanish. Since the singularities of $\st_{\gamma Z}$ and $\st_{21}$ are identical, the cross section to produce $\gamma Z$ is IR finite smoothly through the threshold. Despite the IR finiteness, the physical interpretation of the cancellation at high energy is a little strange: to produce a $\gamma Z$ from $e^+e^-$ initial states, we must also include initial states with $\gamma Z$ in them. On the other hand, we were not originally interested in $\gamma Z$ final states, but $e^+e^-$ initial states, so a more relevant question is what states must we include along with $e^+e^-$ to make a finite cross section?
  
\subsection{Final states only}
Unitarity implies that the sum over final states only, including forward scattering, is IR finite for any initial state. In this case, $e^+e^- \to X$ summed over all states $X$
coming from cuts of the $e^+e^- \to Z$ diagrams must be finite on their own. We already computed $e^+e^- \to \Z$ and $e^+ e^- \to \Z \gamma$, and they did not cancel by themselves. We cannot add additional photons in the initial state, but we must also consider $e^+e^-$ forward scattering with $\gamma Z$ intermediate states. Adding the 2 box diagrams gives:
\begin{multline}
\begin{gathered}
\begin{tikzpicture}
 \node at (0,0) {
\resizebox{30mm}{!}{
     \fmfframe(0,0)(0,0){
\begin{fmfgraph*}(120,100)
 \fmfleft{L1,L2}
        \fmfright{R1,R2}
        \fmf{fermion}{v1,L1}
        \fmf{dashes}{v1,v2}
        \fmf{fermion}{R2,v4}
        \fmf{fermion}{v2,R1}
        \fmf{photon}{v4,v3}
        \fmf{fermion}{L2,v3}
        \fmf{fermion}{v3,v1}
        \fmf{fermion}{v4,v2}
\end{fmfgraph*}
}}};
\draw[dashed,red,thick] (-0.7,1) -- (-0.7,-1);
\end{tikzpicture}
\end{gathered}
+
\begin{gathered}
\begin{tikzpicture}
 \node at (0,0) {
\resizebox{30mm}{!}{
     \fmfframe(0,0)(0,0){
\begin{fmfgraph*}(120,100)
 \fmfleft{L1,L2}
        \fmfright{R1,R2}
        \fmf{fermion}{v1,L1}
        \fmf{phantom}{v1,v2}
        \fmf{fermion}{v2,R1}
        \fmf{fermion}{R2,v4}
        \fmf{phantom}{v4,v3}
        \fmf{fermion}{L2,v3}
        \fmf{fermion}{v3,v1}
        \fmf{fermion}{v4,v2}
        \fmffreeze
        \fmf{dashes}{v1,v4a}
        \fmf{phantom,tension=2}{v4a,v4b}
        \fmf{dashes}{v4b,v4}
\fmffreeze
        \fmf{photon}{v3,v2}
\end{fmfgraph*}
}}};
\draw[dashed,red,thick] (-0.7,1) -- (-0.7,-1);
\end{tikzpicture}
\end{gathered}
%+
%\begin{gathered}
%\begin{tikzpicture}
% \node at (0,0) {
%\resizebox{30mm}{!}{
%     \fmfframe(0,0)(0,0){
%\begin{fmfgraph*}(120,100)
% \fmfleft{L1,L2}
%        \fmfright{R1,R2}
%        \fmf{phantom}{L1,v1}
%        \fmf{dashes}{v1,v2}
%        \fmf{fermion}{R2,v4}
%        \fmf{phantom}{v2,R1}
%        \fmf{photon}{v4,v3}
%        \fmf{fermion}{L2,v3}
%        \fmf{fermion}{v3,v1}
%        \fmf{fermion}{v4,v2}
%        \fmffreeze
%        \fmf{fermion}{v2,L1}
%        \fmf{fermion}{v1,R1}
%\end{fmfgraph*}
%}}};
%\draw[dashed,red,thick] (-0.7,1) -- (-0.7,-1);
%\end{tikzpicture}
%\end{gathered}
+ \text{c.c.}
\\ 
=\sigma_0^d \frac{e^2}{\pi^2} \Gm
\left\{
 \delta(1-z) \left(-\frac{1}{4\eps^2} + \frac{\pi^2}{16}\right)
 + 
 \frac{1+z^2}{4}
 \left(
 \frac{1}{\eps}
  \left[ \frac{1}{1-z} \right]_+
- 2 
  \left[ \frac{\ln (1-z)}{1-z} \right]_+ 
  \right)
  \right\}
  \label{boxes3}
 \end{multline}
These contributions exactly cancel the real emission graphs in Eq.~\eqref{REeeZ}.  

We saw that the IR divergences in the $e^+e^- \to Z$ 1-loop amplitude where canceled by real absorption graphs. To cancel these divergences without absorption graphs we need a different set of forward scattering diagrams, namely those containing the troublesome loop. We find in this case:
\begin{multline}
\begin{gathered}
\begin{tikzpicture}
 \node at (0,0) {
\resizebox{25mm}{!}{
     \fmfframe(0,0)(0,0){
\begin{fmfgraph*}(100,80)
        \fmfleft{L1,L2}
        \fmfright{R1,R2}
        \fmf{phantom}{L1,v1}
        \fmf{phantom}{v1,L2}
        \fmf{fermion,tension=2}{R2,v2}
        \fmf{fermion,tension=2}{v2,R1}
        \fmf{dashes,tension=3}{v1,v2}
        \fmffreeze
        \fmf{fermion}{L1,v3}
        \fmf{fermion,tension=1}{v3,v1}
        \fmf{fermion,tension=1}{v1,v4}
        \fmf{fermion}{v4,L2}
        \fmffreeze
        \fmf{photon}{v3,v4}
\end{fmfgraph*}
}}};
\draw[dashed,red,thick] (-1,1) -- (-1,-1);
\end{tikzpicture}
\end{gathered}
+
\begin{gathered}
\begin{tikzpicture}
 \node at (0,0) {
\resizebox{25mm}{!}{
     \fmfframe(0,0)(0,0){
\begin{fmfgraph*}(100,80)
        \fmfleft{L1,L2}
        \fmfright{R1,R2}
        \fmf{phantom}{R1,v2}
        \fmf{phantom}{v2,R2}
        \fmf{fermion,tension=2}{L2,v1}
        \fmf{fermion,tension=2}{v1,L1}
        \fmf{dashes,tension=3}{v1,v2}
        \fmffreeze
        \fmf{fermion}{R1,v3}
        \fmf{fermion,tension=1}{v3,v2}
        \fmf{fermion,tension=1}{v2,v4}
        \fmf{fermion}{v4,R2}
        \fmffreeze
        \fmf{photon}{v3,v4}
\end{fmfgraph*}
}}};
\draw[dashed,red,thick] (-1,1) -- (-1,-1);
\end{tikzpicture}
\end{gathered}
+ \text{c.c.}
= 
\sigma _ { 0 }^d \frac { e ^ { 2 } } { \pi ^ { 2 } }\Gm \\
\times 
\left\{ \delta ( 1 - z ) \left( \frac { 1 } { 4\eps^2 } + \frac { 3 } { 8\eps } + 1 - \frac { 7 \pi ^ { 2 } } { 48 } \right) + \frac { 1 - z } { ( 1 - z ) ^ { 2 } + \hat{\Gamma}_z^2 } \left( -\frac{ 1 } { 4 \epsilon } - \frac{ 3 } { 8 } \right) \right\}
\end{multline}
where $\hat{\Gamma}_z \equiv \frac{\Gamma_z m_z}{Q^2}$. There is also an additional cut to this diagram, representing a non-forward scattering $e^+e^- \to  e^+ e^-$ contribution:
\be
\begin{gathered}
\begin{tikzpicture}
 \node at (0,0) {
\resizebox{25mm}{!}{
     \fmfframe(0,0)(0,0){
\begin{fmfgraph*}(100,80)
        \fmfleft{L1,L2}
        \fmfright{R1,R2}
        \fmf{phantom}{L1,v1}
        \fmf{phantom}{v1,L2}
        \fmf{fermion,tension=2}{R2,v2}
        \fmf{fermion,tension=2}{v2,R1}
        \fmf{dashes,tension=2}{v1,v2}
        \fmffreeze
        \fmf{fermion,tension=2}{v3,L1}
        \fmf{fermion}{v1,v3}
        \fmf{fermion}{v4,v1}
        \fmf{fermion,tension=2}{L2,v4}
        \fmffreeze
        \fmf{photon}{v3,v4}
\end{fmfgraph*}
}}};
\draw[dashed,red,thick] (-0.7,1) -- (-0.7,-1);
\end{tikzpicture}
\end{gathered}
+ \text{c.c.}
= 
\sigma _ { 0 }^d \frac { e ^ { 2 } } {  \pi ^ { 2 } } \Gm 
\left\{ \frac { 1 - z } { ( 1 - z ) ^ { 2 } + \hat{\Gamma}_z^2 } \left( \frac{ 1 } { 4 \epsilon } + \frac{ 3 } { 8 } \right)  \right\}
\ee
The sum of these graphs cancel the cut graphs in Eq.~\eqref{eeZloop}.

Although the cancellation confirms the general theorem from Sec.~\ref{sec:fsronly}, it is still a bit surprising: to cancel the infrared singularities in $e^+e^- \to \Z$ + photons we must include states without a $\Z$ in them, namely $e^+e^- \to e^+e^-$. One way to explain this observation is that a $Z$ boson can mix with an $e^+e^-$ pair, so the two states are not distinguishable. Actually, this case has extra complications over other QED processes because the $Z$ is massive and unstable (as it must be if it can be produced by massless particles). If the $Z$ were massless, like the photon, the disconnected diagrams with $Z\to e^+e^-$ would not be allowed. We thus turn next to pure QED processes, Compton scattering and light-by-light scattering, to further explore the physics of infrared finiteness and forward scattering.

%%%%%%%%%%%%%%%%%%%%%%%%%%%%%
%%%%%%%%%%%%%. COMPTON
%%%%%%%%%%%%%%%%%%%%%%%%%%%%%
\section{Compton scattering \label{sec:compton}}
Compton scattering is a simple example where where the KLN theorem, summing over degenerate initial and final states, but not including forward scattering, fails. At leading order, Compton scattering has an $s$- and $t$-channel contribution:
\be
\st_{{\color{purple} \text{C}}} =
\left| 
\begin{gathered}
\begin{tikzpicture}
 \node at (-0.05,0) {
\resizebox{20mm}{!}{
     \fmfframe(0,0)(0,0){
\begin{fmfgraph*}(80,80)
       \fmfleft{i1,i2}
        \fmfright{o1,o2}
        \fmf{photon}{i1,v1}
        \fmf{fermion}{v2,v1}
        \fmf{fermion}{v1,o1}
        \fmf{fermion}{i2,v2}
         \fmf{photon}{v2,o2}
\end{fmfgraph*}
}}};
\end{tikzpicture}
\end{gathered}
+
\begin{gathered}
\begin{tikzpicture}
 \node at (-0.05,0) {
\resizebox{20mm}{!}{
     \fmfframe(0,0)(0,0){
\begin{fmfgraph*}(80,80)
       \fmfleft{i1,i2}
        \fmfright{o1,o2}
        \fmf{photon}{i1,v1}
        \fmf{fermion}{v1,v2}
        \fmf{fermion}{v2,o1}
        \fmf{fermion}{i2,v1}
         \fmf{photon}{v2,o2}
         \end{fmfgraph*}
}}};
\end{tikzpicture}
\end{gathered}
\right|^2
\ee
The $s$-channel graph makes a non-singular contribution at finite $Q$, but the $t$-channel graph has a pole.  Regulating the divergence in $d$ dimensions and working in Feynman gauge we find that the t-channel contribution is
\be
\st^{t}_{{\color{purple} \text{C}}} = \frac{e^4}{\pi Q^2}\Gm \left\{ -\frac{1}{2\eps} + 1 \right\}
\ee
with $\Gm = \left( \frac { 4 \pi e ^ { - \gamma_E } \mu ^ { 4 } } { Q ^ { 2 } } \right) ^ { \frac { 4 - d } { 2 } }$.

What could cancel this singularity? We cannot dress the initial or final state electron with additional photons, as the contribution would be higher order in $e$. 
Because of unitarity, and the proof of cancellation, we can find the answer by simply drawing all possible cuts:
\be
\begin{gathered}
\begin{tikzpicture}
 \node at (0,0) {
\resizebox{40mm}{!}{
     \fmfframe(0,0)(0,0){
\begin{fmfgraph*}(150,100)
 \fmfleft{i1,i2}
        \fmfright{o1,o2}
        \fmf{photon}{i1,v1}
        \fmf{fermion}{v1,v2}
        \fmf{photon}{v2,o1}
        \fmf{fermion}{v4,o2}
        \fmf{photon}{v4,v3}
        \fmf{fermion}{i2,v3}
        \fmf{fermion}{v3,v1}
        \fmf{fermion}{v2,v4}
\end{fmfgraph*}
}}};
\draw[dashed,purple,thick] (0,2) -- (0,-2);
\node at (0+0.2,2) {\color{purple} C};
\draw[dashed,brown,thick] (-1.5,2) -- (-1.5,-2);
\node at (-1.5-0.2,2) {\color{brown} F};
\draw[dashed,brown,thick] (1.5,2) -- (1.5,-2);
\node at (1.5-0.2,2) {\color{brown} F${}^\star$};
\draw[dashed,orange,thick] plot [smooth] coordinates {(-1.25,2)  (-1.25,0.5) (-0.5,0) (-0.5,-2)};
\node at (-1,2) {\color{orange} D$_1$};
\draw[dashed,red,thick] (-0.5,2) .. controls (0,-2) and (-1.5,1) .. (-1.25,-2);
\node at (-0.3,2) {\color{red} D$_2$};
\end{tikzpicture}
\end{gathered}
\label{ComptonCuts}
\ee
Actually, rather drawing the cuts as lines (or shaded lines), we find it clearest to enumerate the possible cuts as all possible circlings of the vertices, following 't Hooft and Veltman~\cite{tHooft:1973wag,Veltman:1994wz}. For example, 
\be
\begin{gathered}
\begin{tikzpicture}
\node at (-1,0) {{\color{orange} D$_1$} = };
 \node at (0,0) {
\resizebox{20mm}{!}{
     \fmfframe(0,0)(0,0){
\begin{fmfgraph*}(75,75)
 \fmfleft{i1,i2}
        \fmfright{o1,o2}
        \fmf{photon}{i1,v1}
        \fmf{fermion}{v1,v2}
        \fmf{photon}{v2,o1}
        \fmf{fermion}{v4,o2}
        \fmf{photon}{v4,v3}
        \fmf{fermion}{i2,v3}
        \fmf{fermion}{v3,v1}
        \fmf{fermion}{v2,v4}
\end{fmfgraph*}
}}};
\draw[orange,thick] (-0.3,0.3) circle (0.1);
\draw[orange,thick] (0.3,0.3) circle (0.1);
\draw[orange,thick] (0.3,-0.3) circle (0.1);
\end{tikzpicture}
\end{gathered}
,
\begin{gathered}
\begin{tikzpicture}
\node at (-1,0) {{\color{purple} C} = };
 \node at (0,0) {
\resizebox{20mm}{!}{
     \fmfframe(0,0)(0,0){
\begin{fmfgraph*}(75,75)
 \fmfleft{i1,i2}
        \fmfright{o1,o2}
        \fmf{photon}{i1,v1}
        \fmf{fermion}{v1,v2}
        \fmf{photon}{v2,o1}
        \fmf{fermion}{v4,o2}
        \fmf{photon}{v4,v3}
        \fmf{fermion}{i2,v3}
        \fmf{fermion}{v3,v1}
        \fmf{fermion}{v2,v4}
\end{fmfgraph*}
}}};
\draw[purple,thick] (0.3,0.3) circle (0.1);
\draw[purple,thick] (0.3,-0.3) circle (0.1);
\end{tikzpicture}
\end{gathered}
,
\begin{gathered}
\begin{tikzpicture}
\node at (-1,0) {{\color{brown} F${}^\star$} = };
 \node at (0,0) {
\resizebox{20mm}{!}{
     \fmfframe(0,0)(0,0){
\begin{fmfgraph*}(75,75)
 \fmfleft{i1,i2}
        \fmfright{o1,o2}
        \fmf{photon}{i1,v1}
        \fmf{fermion}{v1,v2}
        \fmf{photon}{v2,o1}
        \fmf{fermion}{v4,o2}
        \fmf{photon}{v4,v3}
        \fmf{fermion}{i2,v3}
        \fmf{fermion}{v3,v1}
        \fmf{fermion}{v2,v4}
\end{fmfgraph*}
}}};
\end{tikzpicture}
\end{gathered}
,
\quad\text{etc.}
\ee
Lines going from uncircled vertices to uncircled vertices get a $+ i\eps$, lines going from circled to circled get a $ - i \eps$ and lines going from uncircled to circled are
cut, so they get $\delta$ functions. Although not explained explicitly in~\cite{tHooft:1973wag,Veltman:1994wz}, an incoming line connecting to a circled vertex or an outgoing line connecting to an uncircled vertex, as in diagram {\color{orange} D${}_1$} gives a disconnected line. In this way, we see that there are 16 possible cuts. Most of these vanish. Indeed, if a connected set of circled vertices attaches only to incoming lines or a connected set of uncircled vertices attaches only to outgoing lines, the graph vanishes by energy conservation. Thus,
\be
\begin{gathered}
\begin{tikzpicture}
 \node at (0,0) {
\resizebox{20mm}{!}{
     \fmfframe(0,0)(0,0){
\begin{fmfgraph*}(75,75)
 \fmfleft{i1,i2}
        \fmfright{o1,o2}
        \fmf{photon}{i1,v1}
        \fmf{fermion}{v1,v2}
        \fmf{photon}{v2,o1}
        \fmf{fermion}{v4,o2}
        \fmf{photon}{v4,v3}
        \fmf{fermion}{i2,v3}
        \fmf{fermion}{v3,v1}
        \fmf{fermion}{v2,v4}
\end{fmfgraph*}
}}};
\draw[red,thick] (-0.3,-0.3) circle (0.1);
\draw[red,thick] (-0.3,0.3) circle (0.1);
\draw[red,thick] (0.3,-0.3) circle (0.1);
\end{tikzpicture}
\end{gathered}
=0
,
\begin{gathered}
\begin{tikzpicture}
 \node at (0,0) {
\resizebox{20mm}{!}{
     \fmfframe(0,0)(0,0){
\begin{fmfgraph*}(75,75)
 \fmfleft{i1,i2}
        \fmfright{o1,o2}
        \fmf{photon}{i1,v1}
        \fmf{fermion}{v1,v2}
        \fmf{photon}{v2,o1}
        \fmf{fermion}{v4,o2}
        \fmf{photon}{v4,v3}
        \fmf{fermion}{i2,v3}
        \fmf{fermion}{v3,v1}
        \fmf{fermion}{v2,v4}
\end{fmfgraph*}
}}};
\draw[red,thick] (-0.3,0.3) circle (0.1);
\draw[red,thick] (0.3,-0.3) circle (0.1);
\end{tikzpicture}
\end{gathered}
=0
,
\begin{gathered}
\begin{tikzpicture}
 \node at (0,0) {
\resizebox{20mm}{!}{
     \fmfframe(0,0)(0,0){
\begin{fmfgraph*}(75,75)
 \fmfleft{i1,i2}
        \fmfright{o1,o2}
        \fmf{photon}{i1,v1}
        \fmf{fermion}{v1,v2}
        \fmf{photon}{v2,o1}
        \fmf{fermion}{v4,o2}
        \fmf{photon}{v4,v3}
        \fmf{fermion}{i2,v3}
        \fmf{fermion}{v3,v1}
        \fmf{fermion}{v2,v4}
\end{fmfgraph*}
}}};
\draw[red,thick] (-0.3,-0.3) circle (0.1);
\end{tikzpicture}
\end{gathered}
=0
,
\quad\text{etc.}
\ee
Enumerating the cuts through circled vertices is particularly helpful for disconnected diagrams, like the crossed-box graphs in Eq.~\eqref{boxes3}, where drawing lines through the graph is ambiguous. For example,
\be
\begin{gathered}
\begin{tikzpicture}
 \node at (0,0) {
\resizebox{20mm}{!}{
     \fmfframe(0,0)(0,0){
\begin{fmfgraph*}(75,75)
        \fmfleft{L1,L2}
        \fmfright{R1,R2}
        \fmf{fermion,tension=2}{L2,v2}
        \fmf{fermion,tension=2}{w2,R2}
        \fmf{photon,tension=2}{L1,v1}
        \fmf{photon,tension=2}{w1,R1}
        \fmf{phantom}{v1,v2}
        \fmf{photon}{v2,w2}
        \fmf{phantom}{w2,w1}
        \fmf{fermion}{w1,v1}
        \fmffreeze
        \fmf{fermion}{v2,w1}
        \fmf{fermion}{v1,v4a}
        \fmf{phantom,tension=1}{v4a,v4b}
        \fmf{plain}{v4b,w2}
\end{fmfgraph*}
}}};
\draw[red,thick] (-0.44,0.44) circle (0.1);
\draw[red,thick] (0.44,-0.44) circle (0.1);
\end{tikzpicture}
\end{gathered}
=
\begin{gathered}
\begin{tikzpicture}
 \node at (0,0) {
\resizebox{20mm}{!}{
\fmfframe(0,0)(0,0){
\begin{fmfgraph*}(80,80)
\fmfstraight
\fmfleft{L1,L1b,L1c,L2,L3,L4,L5,L6}
\fmfright{R1,R1b,R1c,R2,R3,R4,R5,R6}
\fmf{phantom}{R6,v2}
\fmf{phantom}{v2,R2}
\fmf{photon,tension=2}{v2,L4}
\fmf{fermion}{L1,R1}
\fmffreeze
\fmf{fermion}{R2,v2}
\fmf{fermion}{v2,v1}
\fmf{fermion}{v1,R6}
\fmffreeze
\fmf{photon}{v1,R4}
\end{fmfgraph*}
}}};
\end{tikzpicture}
\end{gathered}
\begin{gathered}
\begin{tikzpicture}
\draw[dashed,red,thick] (-1.1,0.8) -- (-1.1,-0.8);
 \node at (0,0) {
\resizebox{20mm}{!}{
\fmfframe(0,0)(0,0){
\begin{fmfgraph*}(80,80)
\fmfstraight
\fmfleft{L2,L3,L4,L5,L6,L7}
\fmfright{R2,R3,R4,R5,R6,R7}
\fmf{phantom}{L2,v2}
\fmf{phantom}{v2,L6}
\fmf{photon,tension=2}{v2,R4}
\fmf{fermion}{L7,R7}
\fmffreeze
\fmf{photon}{L6,v1}
\fmf{fermion}{v2,v1}
\fmf{fermion}{L2,v2}
\fmffreeze
\fmf{fermion}{v1,L4}
\end{fmfgraph*}
}}};
\end{tikzpicture}
\end{gathered}
\ee
This is the connected component of the interference between the disconnected and connected graph (see discussion in Sec. \ref{sec:ifsum}). It vanishes for Compton scattering since the photon is massless even though the equivalent topology for $\gamma Z\to e^+e^- Z$ in Eq.~\eqref{twotoone}, does not vanish.

Now let us look at some of the graphs. Cut  {\color{orange} D${}_1$} is
\be
\begin{gathered}
\begin{tikzpicture}
 \node at (0,0) {
\resizebox{30mm}{!}{
     \fmfframe(0,0)(0,0){
\begin{fmfgraph*}(120,60)
\fmfleft{i1,i2}
\fmfright{o1,o2}
    \fmf{fermion,tension=0.3}{i2,v1}
    \fmf{photon,tension=1.5}{i1,v9}
    \fmf{phantom,tension=2}{v1,v4}
    \fmf{fermion}{v4,v10,v5}
    \fmf{fermion,tension=1.5}{v7,o2}
    \fmf{photon,tension=2}{v10,v7}
    \fmf{fermion,tension=0.5}{v2,v9,v3}
    \fmf{phantom,tension=2}{v2,v5}
    \fmf{phantom,tension=2}{v3,v6}
    \fmf{fermion,tension=0.5}{v6,v8}
    \fmf{fermion}{v8,v7}
    \fmf{photon}{v8,o1}
    \end{fmfgraph*}
}}};
%  \draw[style=help lines] (-2,-2) grid (2,2); 
\draw[dashed,orange,thick] (-0,1.2) -- (0.1,-1.2);
\node at (-0.3,1.5) {\color{orange} D$_1$};
\end{tikzpicture}
\end{gathered}
\ee
This diagram is the product of a disconnected graph for $e^- \to e^-$ and  $\gamma \to e^+e^-$ and a connected graph for  $e^- \gamma \to e^- e^+ e^-$. The disconnected part only has support if the $e^+e^-$ pair is collinear to the incoming photon. In this phase space region, which is a set of measure zero over all of phase space, the connected part is non-singular and thus the product vanishes when integrated over phase space. (The disconnected diagrams in Eq.~\eqref{onetoone} had an $e^+e^- \to Z$ component which did not vanish because the $Z$ boson is massive.) In is not hard to see that for Compton scattering at this order, all the cuts giving disconnected graphs are exactly 0. 

Thus the only remaining  contribution to cancel the divergence in the $\st_{{\color{purple} \text{C}}}$ 
is forward scattering. The forward scattering contribution, labeled {\color{brown} F} in Eq.~\eqref{ComptonCuts} is the interference of the forward scattering non-interacting diagram and the box
\be
\st_{\color{brown} \text{F}}=
\hspace{-0.5cm}
\begin{gathered}
\begin{tikzpicture}
 \node at (0,0) {
\resizebox{40mm}{!}{
     \fmfframe(0,0)(0,0){
\begin{fmfgraph*}(150,80)
\fmfleft{i1,i2}
\fmfright{o1,o2}
\fmf{photon}{i1,w2}
\fmf{phantom,tension=2}{w2,w3}
\fmf{photon}{w3,v1}
\fmf{fermion}{i2,x1}
\fmf{phantom,tension=2}{x1,x2}
\fmf{fermion}{x2,v3}
\fmf{fermion}{v1,v2}
\fmf{photon}{v2,o1}
\fmf{fermion}{v4,o2}
\fmf{photon}{v4,v3}
\fmf{fermion,tension=0.5}{v3,v1}
\fmf{fermion}{v2,v4}
\end{fmfgraph*}
}}};
\draw[dashed,brown,thick] (-0.8,1) -- (-0.8,-1);
\node at (-0.3,1) {\color{brown} F};
\end{tikzpicture}
\end{gathered}
=
\frac{e^4}{\pi Q^2}\Gm \left\{ \frac { i } { 4 \pi \eps^2 } + \frac{ - 2 i + \pi }{ 4 \pi \eps } - \frac { 1 } { 2 } + \frac{ i } { 4 \pi } - \frac { 7 i \pi } { 48 }  \right\}
\ee
In this contribution, the tree-level part only has support for forward scattering and the loop however is singular at $t=0$. Their product is integrable in $d$ dimensions, leading to the above result. 

Adding to this the cut graph ${{\color{brown} \text{F}^\star}}$ , which is the complex conjugate to ${{\color{brown} \text{F}}}$ gives
\be
\sigma_{\color{brown} \text{F}}  + \sigma_{\color{brown} \text{F}^\star} = \frac{e^4}{\pi Q^2}\Gm \left\{ \frac{1}{2\eps} - 1 \right\}
\ee
which exactly cancels the tree-level cross section, as expected. 

On the one hand, this result should not come as a surprise. It is guaranteed by unitarity. However, note that the Compton diagrams had a singularity in the Bjorken $x=1$ region, where the entire momentum of the incoming electron is transferred to the outgoing photon. In contrast, the forward scattering contribution is at $x=0$, where the momentum of the electron stays with the electron.  Thus in the two cancelling contributions the hard electron is going off in entirely different directions.
 It seems like the question of whether a hard particle is an electron or photon should be physical. We find that instead, only the cross section for an hard electron {\it or} hard photon is finite.

One way to understand why electrons and photons are effectively indistinguishable at high energy is that when the electron is massless, there is no energetic penalty to produce additional $e^+e^-$ pairs from the vacuum. Thus, the state with a photon and electron can mix with one where a soft positron is created in the  hard electron's direction, neutralizing its charge to produce a photon, and a soft electron is created in the hard photon's direction.  Thus a hard electron going left and a hard photon going right can mix with the state of a hard electron going right and hard photon going left. Such mixing are exactly the degeneracies that must be summed over in the KLN theorem to get a finite result. 

One could object to the reasoning here because the electron is in fact massive. Indeed, there are no massless particles in nature with nonzero electric charge, and such particles may not even be consistent (although gluons are, of course, massless particles charged under a different force). The point, however, is not to envision some fictitious theory with massless electrons.
Rather, we want to understand when and how large contributions to the cross section are counterbalanced by superficially distinguishable processes.
If the electron had a small mass, the collinear divergence would be regulated. We would then find the rate for producing a hard electron scales like $e^4 \ln \frac{Q}{m_e}$ at large $\frac{Q}{m_e}$ and the rate for producing a hard forward photon scales like $1-e^4\ln \frac{Q}{m_e}$. 
Thus at high energy, when the logarithms become large, the two contributions should be added to restore perturbativity. In terms of the physical picture, when the center of mass energy of the collision becomes high enough, the energetic penalty to produce $e^+e^-$ pairs takes a negligible amount of the total energy. In this way, a hard electron and hard photon become indistinguishable and their cross sections must be combined, according to the same logic as when $m_e =0$. 

\section{$\gamma \gamma \to X$ \label{sec:photon}}
Perhaps the most powerful example of the failure of the initial-and-final-state sum picture is for light-by-light scattering -- it is hard to argue that a photon is not a well-defined asymptotic state and that additional initial state particles must be added in $\gamma \gamma \to \gamma \gamma$. Let us consider then the process $\gamma \gamma \to e^+e^-$. The total cross section for this process is IR divergent due to the forward scattering region:
\be
\left|
\begin{gathered}
\begin{tikzpicture}
 \node at (-0.05,0) {
\resizebox{20mm}{!}{
     \fmfframe(0,0)(0,0){
\begin{fmfgraph*}(80,80)
\fmfleft{L1,L2}
\fmfright{R1,R2}
\fmf{photon}{L1,v1}
\fmf{photon}{L2,v2}
\fmf{fermion}{R2,v2}
\fmf{fermion}{v2,v1}
\fmf{fermion}{v1,R1}
\end{fmfgraph*}
}}};
\end{tikzpicture}
\end{gathered}
+
\begin{gathered}
\begin{tikzpicture}
 \node at (-0.05,0) {
\resizebox{20mm}{!}{
     \fmfframe(0,0)(0,0){
\begin{fmfgraph*}(80,80)
\fmfleft{L1,L2}
\fmfright{R1,Ra,Rb,Rc,R2}
\fmf{photon}{L1,v1}
\fmf{photon}{L2,v2}
\fmf{phantom}{R2,v2}
\fmf{fermion}{v1,v2}
\fmf{phantom}{v1,R1}
\fmffreeze
\fmf{phantom}{Rc,v1}
\fmf{fermion}{v2,Ra}
\fmffreeze
\fmf{fermion}{Rc,va}
\fmf{phantom,tension=2}{va,vb}
\fmf{plain}{vb,v1}
\end{fmfgraph*}
}}};
\end{tikzpicture}
\end{gathered}
\right|^2
=
\frac { e ^ { 4 } } { \pi Q ^ { 2 } }\Gm \left\{ - \frac { 1 } { \eps } + 1 \right\}
\ee
with $\Gm = \left(\frac {4 \pi e ^ { - \gamma_E } \mu ^ { 4 } } { Q ^ { 2 } } \right) ^ { \frac { 4 - d } { 2 } }$. As with Compton scattering, since the divergence is at tree-level, there can be no loop or bremsstrahlung contributions to cancel this singularity. Instead, the singularity is canceled by the forward scattering amplitude
\be
\begin{gathered}
\begin{tikzpicture}
 \node at (0,0) {
\resizebox{30mm}{!}{
     \fmfframe(0,0)(0,0){
\begin{fmfgraph*}(120,100)
 \fmfleft{L1,L2}
        \fmfright{R1,R2}
        \fmf{photon}{L1,v1}
        \fmf{fermion}{v1,v2}
        \fmf{photon}{R1,v2}
        \fmf{photon}{R2,v4}
        \fmf{fermion}{v4,v3}
        \fmf{photon}{L2,v3}
        \fmf{fermion}{v3,v1}
        \fmf{fermion}{v4,v2}
\end{fmfgraph*}
}}};
\draw[dashed,red,thick] (-0.7,1) -- (-0.7,-1);
\end{tikzpicture}
\end{gathered}
+ \text{crossings} + \text{c.c.} =  
\frac { e ^ { 4 } } { \pi Q ^ { 2 } } \Gm \left\{  \frac { 1 } { \eps } - 1 \right\}
\ee 
The sum of these is exactly zero, as expected by unitarity. Note that for this process, as for Compton scattering, all the diagrams with $1\to 2$ disconnected pieces vanish exactly since all the particles are massless. 

We have found something shocking: the total rate for photons to annihilate into charged particles is undefined. Similarly, the total rate for photons to annihilate into photons is undefined. Only the total cross section including photon final states and $e^+e^-$ final states is IR finite. 

\section{Summary and Conclusions \label{sec:conc}}
This paper explores the question of which cross sections must be summed along with the cross section for a given process to produce an infrared finite result. Some of the main results of this paper are that

\begin{enumerate}
\item One never needs to sum over initial {\it and} and final states to achieve IR finiteness, in contrast to expectations from the KLN theorem.
\item IR finiteness often requires the inclusion of forward scattering and the interference between disconnected and connected Feynman diagrams. 
\item In QED with massless electrons, $e^+e^- \to Z$ can be made IR finite at  the first nontrivial order by including
\begin{enumerate}
\item outgoing photons and $e^+e^- \to e^+e^-$,
\item incoming photons and $Z\to Z$, or
\item an infinite number of processes dressing $e^+e^- \to Z$ with additional incoming or outgoing photons. Summing all the contributions, the cross section with fixed initial-state jet masses is convergent.
\end{enumerate}
\item In $e^- \gamma \to e^- \gamma$, the tree-level IR divergence from the region with the outgoing $\gamma$ collinear to the incoming $e^-$ is canceled by the region with the outgoing $\gamma$ collinear to the incoming $\gamma$. 
\item The IR divergence in $\gamma\gamma \to \gamma \gamma$ scattering is cancelled by $\gamma \gamma \to e^+e^-$.
\end{enumerate}

The first point is perhaps the most important observation in this paper.  Although the KLN theorem instructs us to sum over degenerate initial and final states to produce an infrared finite cross section, in fact only the sum over initial {\it or} final states is necessary. Moreover, finiteness is only guaranteed if the forward scattering contribution is included.

In some cases, the forward scattering contribution is infrared finite on its own. An important example is $Z\to Z$. Its finiteness allows the rate for $Z\to e^+e^- +$ photons to be finite, or $Z \to \text{hadrons}$ to be finite in QCD. That the infrared singularities (in particular the collinear singularities associated with massless electrons or quarks)  cancel is in a sense an accident of the simplicity of the $Z\to Z$ amplitude. For most other processes, forward scattering is singular and must be included for infrared finiteness. For example, $Z\gamma \to Z\gamma$ at 1-loop is IR finite only in a small kinematic window but otherwise divergent.

Although one may sum only over final states, it may be important to consider initial state sums in some contexts. The example we studied in detail here was $e^+e^-   \to Z $. For this process, one can add photons to the initial state and the cross section will be finite, as the process is then the exact crossing of $Z\to e^+e^-$. However, for $e^+e^- \to Z$, one cannot prevent the electrons from radiating photons into the final state. These $e^+e^- \to Z \gamma$ processes are infrared divergent, with their infrared divergences
canceled in turn  by additional diagrams with disconnected photons. We understood this cancellation as a initial-state sum cancellation because the $Z\gamma$ forward scattering amplitude is infrared finite when the center-of-mass energy is close to $m_Z$.  Moreover, we found an infinite number of diagrams contributing at next-to-leading order in perturbation theory. Summing these diagrams, all the infrared divergences cancel. The infinite sum over the finite parts of all the diagrams appears to be convergent, although it is not clear how to interpret the result as hard wide-angle initial state photons are included. In Appendix~\ref{app:hemispheres}, we refined the calculation to initial-state jet masses, to eliminate the hard wide-angle photons, and still found convergence. It will certainly be interesting to consider connecting these infinite sums to experimental observables, as this is an example where an initial and final state sum gives a non-trivial result. Two reasons this may be challenging are that 1) the infinite sum over the finite parts is convergent but not absolutely convergent, so the result depends on how the terms are ordered and 2) it is not clear if the cancellation will hold at higher orders, or in more complicated theories like QCD. 

An alternative to summing over initial and final states is to  sum over just final states, but to  include also the $e^+e^- \to Z \to e^+e^-$ forward scattering contribution. This contribution is IR divergent and cancels the IR divergence of the $e^+e^- \to Z$ virtual corrections as well as the $e^+e^- \to Z \gamma$ bremsstrahlung graphs. Thus, the forward scattering in this case achieves the same cancellation as the multiple initial and final state photons did, but avoids having to include disconnected diagrams and perform an infinite sum. On the other hand, when forward scattering is included, the total cross section is exactly zero at this order (as required by unitarity). 

Additional insight came from examining Compton scattering. In Compton scattering, the total cross section with massless electrons is IR divergent. With an small electron mass, the total cross section diverges as $\sigma \sim \frac{e^4}{32\pi Q^2} \ln \frac{Q^2}{m_e^2}$. The singularity is from the kinematic region where the outgoing photon is collinear to the incoming electron. The large logarithm is canceled by the process $e^- \gamma \to e^- \gamma$ at 1-loop interfered with the disconnected forward scattering amplitude, so the outgoing photon is collinear to the incoming {\it photon}. This says that if a cosmic ray electron comes in at ultra-high energy, and scatters off a photon in the atmosphere one should not be able to distinguish a high-energy photon coming towards us from a high-energy electron. Only the sum of the two cross sections is IR finite (or free of large logarithms at high energy). We presented a physical justification for the indistinguishability: at very high energy, there is negligible energetic cost to the photon converting to an $e^+e^-$ pair. If the positron produced is soft and the electron goes in the photon direction, then effectively the photon has transformed into an electron. 
From a practical point of view, since the electron is in fact massive and clearly distinguishable from a photon when it is slow, the criterion for distinguishability must depend on some experimental resolution to identifying a conversion or charged tracks.

More broadly, we must question when the charge flowing into a certain direction is observable or only the net (global) charge. It seems that in addition to experimental limits on the energy and angles that can be resolved, there must also be an experimental limit on how well the momentum of a charged particle can be measured. That is, the notion of infrared-and-collinear safety might need to be extended to a restriction on charge measurement when massless initial states are involved (of if large logarithms are to cancel when initial state charged particles have mass).

Part of the reason we began investigating the KLN theorem was to gain a handle on the intricate subject of asymptotic states and the $S$-matrix.
In particular, there are proposals that the $S$-matrix might be rendered IR finite if initial and final states are dressed as coherent states. While the original proposals focused on QED with massive electrons~\cite{Chung:1965zza,Kibble:1969ip,Kulish:1970ut,Zwanziger:1973if} there have been extensions to the cases with massless charged particles~\cite{Contopanagos:1991yb,DelDuca:1989jt,Forde:2003jt}, QCD~\cite{Curci:1979bg,Giavarini:1987ts} and gravity~\cite{Ware:2013zja}. While this coherent state approach is intuitively appealing -- it certainly makes sense in the context of the $e^+e^- + \text{photons} \to Z + \text{photons}$ case we studied here -- our observations indicate that the cancellations observed may be accidental. For example, we discussed photon scattering in Sec.~\ref{sec:photon}. We showed there that the cross section for $\gamma \gamma \to $ photons is infinite in a theory with massless electrons. This IR divergence is canceled by the process $\gamma \gamma \to e^+e^-$. In a coherent state approach, one would attempt to achieve the cancellation at the amplitude level, but this would involve dressing the photons with electrons.  While such a dressing is not inconceivable, it deviates from the
the Faddeev-Kulish idea that the IR divergences originate from the long-range interactions in the Hamiltonian. If it is possible to dress states so that the $S$-matrix is finite, the integrals involved in the dressing are likely to be closely related to integrals involved in achieving finite cross sections, like those we have studied here.

Finally, it is worth ruminating on how to connect infrared finiteness forward scattering to experimentally testable predictions. They key may be to understand better
the initial state sums. Indeed, although we have shown that one {\it can} achieve IR finiteness with just a final state sum, it is not clear that this is the most physical way to proceed. 
 A case in point is massive-electron QED, where the Bloch-Nordsieck theorem holds. In QED a final-state sum is sufficient in any process, such as for $e^+e^- \to Z + \text{photons}$. However, our analysis in Section~\ref{sec:eeZ} demonstrated that combining final state emission of $e^+e^- \to Z\gamma$ with the virtual $e^+e^- \to Z$ loop is morally equivocal:\footnote{ 
A counterargument based on spacetime symmetries~\cite{Strominger:2017zoo} can be found in~\cite{Kapec:2017tkm}.}
 these contributions do not come from cuts of the same graph and their cancellation is accidental (a consequence of Abelian exponentiation). 
 Indeed, in the massless-electron case, the cancellation does not work without also including $e^+e^- \to e^+e^-$ forward scattering.
 Alternatively, once can cancel the $e^+e^- \to Z$ loop against $\gamma e^+e^- \to Z$ graphs. Doing so not only cancels the IR divergences, but also the large logarithms of $\frac{m_e}{Q}$. Thus we actually have 3 different ways to compute $e^+e^-$ annihilation:
 1) with a final state sum, \`a la Bloch-Nordseick, whereby a large logarithm results 2) with a final state sum, including $e^+e^- \to e^+e^-$ whereby the inclusive cross section is zero or 3) with an initial state sum, where a finite cross section with no large logarithms results. Of these, the 3$^{\text{rd}}$ option may be the most appealing. However, to actually connect initial-state jets and disconnected diagrams to experiment will require understanding initial state sums in greater detail, to higher order, and in a more complicated yet more experimentally accessible theory, QCD.

\section*{Acknowledgements}
The authors would like to thank Martin Beneke, John Collins, Aneesh Manohar, George Sterman and Hua-Xing Zhu for helpful discussions. 

\appendix
\section{On-shell intermediate propagators \label{app:onshell}}
In this appendix, we explain how to compute cut diagrams with on-shell intermediate propagators, as in Eq.~\eqref{onetoone}. Consider the contribution to the total cross section for $\gamma Z$ scattering with a final state $e^+e^-\gamma$, interfered with a disconnected diagram:
\be
\sigma_{11.A}=
\begin{gathered}
\begin{tikzpicture}
\draw[white] (-1,0.8) -- (-1,-0.8);
 \node at (0,0.5) {
\resizebox{20mm}{!}{
     \fmfframe(0,0)(0,0){
\begin{fmfgraph*}(80,60)
\fmfright{L1,y1,y2,L2,x1,x2,L3}
\fmfleft{R1,w1,w2,R2,z1,z2,R3}
\fmf{dashes,tension=3}{z1,v3}
\fmf{phantom}{L3,v3}
\fmf{phantom}{v3,L1}
\fmffreeze
\fmf{fermion}{L3,v3}
\fmf{fermion}{v3,v2}
\fmf{fermion}{v2,v4}
\fmf{photon}{v4,L1}
\fmffreeze
\fmf{photon}{v2,R1}
\fmf{fermion}{v4,y2}
\end{fmfgraph*}
}}};
\end{tikzpicture}
\end{gathered}
\begin{gathered}
\begin{tikzpicture}
\draw[dashed,red,thick] (-1.3,1) -- (-1.3,-0.8);
 \node at (0,-0) {
\resizebox{20mm}{!}{
\fmfframe(0,0)(0,0){
\begin{fmfgraph*}(80,80)
\fmfstraight
\fmfright{L1,L2,L3,L4,L5}
\fmfleft{R1,R2,R3,R4,R5}
\fmf{dashes,tension=2}{L4,v1}
\fmf{fermion}{v1,R5}
\fmf{fermion}{R3,v1}
\fmffreeze
\fmf{photon}{L2,R2}
\end{fmfgraph*}
}}};
\node at (1.3,0.5) {$p_Z$};
\node at (1.3,-0.5) {$k'$};
\node at (-1.1,0.8) {$q$};
\node at (-1.1,0) {$p$};
\node at (-1.1,-0.5) {$k$};
\end{tikzpicture}
\end{gathered}
+\text{c.c.}
\ee
This is the same as the first diagram in Eq.~\eqref{onetoone}, but crossed so the $\gamma Z$ is incoming. We do this to separate this complication of on-shell intermediate states from that of integrating over 3-body initial-state phase space. The spin-summed cut diagram is
\begin{multline}
\sigma_{11.A}= \sigma_0
\int \frac{d^d k}{(2\pi)^d}\frac{d^d q}{(2\pi)^d}  \frac{d^d p}{(2\pi)^d} 2\pi \delta(k^2) \theta(k_0)2\pi\delta(p^2)\theta(p_0) 2\pi \delta(q^2)\theta(q_0)\\
\times (2\pi)^d \delta^d(q+p-p_Z) (2\pi)^{d-1} (2\omega_k) \delta^{d-1}(k' - k)\\
\times \frac{i}{(p+k)^2+i\eps} \frac{i}{(p+k-k')^2+i\eps} \mathrm{Tr}[ \slashed{p}\gamma^\mu (\slashed{p} + \slashed{k})\gamma^\mu (\slashed{p} + \slashed{k} - \slashed{k'})\gamma^\alpha \slashed{q}\gamma^\alpha] + \text{c.c.}
% \int d\Pi {\cal M}_L  {\cal M}_R^\star
\end{multline} 
The $(2\pi)^{d-1} (2\omega_k) \delta^{d-1}(k' - k)= \langle k | k'\rangle$ factor on the second line comes from projecting the incoming photon momentum onto the outgoing photon momentum in the absence of interactions. Integrating over $d^d k d^d q$ causes no problems. But once $k=k'$, the integral reduces to $\delta(p^2)\frac{i}{p^2+i\eps}$, which must be treated carefully. Integrating over all variables other than $p_0$ and $\omega_p = |\vec{p}|$  gives
\begin{multline}
\sigma_{11.A}= \sigma_0 \frac{\Omega_{d-2}\left(- 2 (d-2)^2\right)}{(2\pi)^{d-2}(1-z)} \int_{-\infty}^\infty d p_0 \int_\frac{z}{2}^\frac{1}{2} d \omega_p
\delta(p_0^2 - \omega_p^2)\left[ \frac{1}{p_0^2 - \omega_p^2 + i \eps} + \text{c.c.} \right] \theta(p_0)
 \\
\times 
 \omega_p^{d-3} \left[\frac{[ (1-p_0)^2 - \omega_p^2] [ \omega_p^2 - (p_0 - z)^2]}{(1-z)^2 \omega_p^2}\right]^{\frac{d-4}{2}}
\frac{p_0^4 - 2 p_0 z + z^2 - (1-z) \omega_p^2 + \omega_p^4 - p_0^2(2 \omega_p^2 + z -1)}{2p_0-z} \label{s1form}
\end{multline}
While $p_0^2 - \omega_p^2$ has two roots, the root with $p_0=-\omega_p$ is off the integration contour due to the $\theta(p_0)$ in the integrand. Thus we can drop the $i\eps$ term for the $p_0 + \omega_p$ factor and focus on the singularity at $p_0 =\omega$. For this singularity, it is ciritical to treat the product $\delta(p_0 - \omega_p)\left[ \frac{1}{p_0 - \omega_p + i \eps} + \text{c.c.} \right]$ as a distribution. By taking the derivative
of the relation
$2\pi \delta(x) = \frac{i}{x+i \eps} - \frac{i}{x-i\eps}$ 
we are led to
\be
2\pi \delta'(x) =- i \left( \frac{1}{x+i \eps}\right)^2 + i \left( \frac{i}{x-i\eps}\right)^2
\ee
Thus we can write
\be
\delta(p_0- \omega_p)\left[ \frac{1}{p_0 - \omega_p + i \eps} + \text{c.c.} \right] =
\frac{i}{2\pi}\left( \frac{1}{p_0 - \omega_p + i \eps} \right)^2 -\frac{ 1}{2\pi}\left( \frac{1}{p_0 - \omega_p - i \eps} \right)^2
=- \delta'(p_0 - \omega_p)
\ee
The $\delta'(p_0-\omega_p)$ can then be integrated by parts. So, for a test function $f(p_0)$ we have
\be
\int_0^\infty d p_0 \delta(p_0^2 - \omega_p^2)\left[ \frac{1}{p_0^2 - \omega_p^2 + i \eps} + \text{c.c.} \right] f(p_0)
=\frac{d}{d p_0} \left[ \left(\frac{1}{p_0 + \omega_p}\right)^2 f(p_0)\right]_{p_0 = \omega_p} \label{dprel}
\ee
%Although one can check this relation easily for $f(p_0)=1$ using the residue theorem for double poles, in general it is incorrect  to integrate over $p_0$ without treating the integrand as a distribution. 
Applying this prescription
%the prescription in Eq. \eqref{dprel}
 to Eq.~\eqref{s1form} gives
\be
\sigma_{11.A}= \sigma_0 \Gm \frac{1}{\pi} \left\{\frac{1}{\eps} - 3 - \ln z \right\}
\ee
The same technique is used to compute Eq.~\eqref{onetoone}. 

A highly non-trivial check on this procedure is that the cross section for $\gamma Z \to \gamma e^+e^-$ computed this way exactly cancels the contributions from other $\gamma Z$ final states at the same order in perturbation theory. In particular, the other diagrams, such as the forward scattering loop and the $\gamma Z \to e^+e^-$ process are computed without having singular intermediate propagators.

\section{Initial state masses \label{app:hemispheres}}
In Section~\ref{sec:eeZ} we showed that the cross section for $n \gamma + e^+e^- \to m \gamma + Z$ was IR finite for each $n$, summed over $m$. Because of its finiteness and the possible convergence of the sum over $n$, one might hope to connect the cross section to a physical observable. To do so, the total cross section, inclusive over all possible initial state photons, including hard non-collinear ones, is probably not the most sensible thing to try to measure. To refine the calculation to something closer to physical, we consider instead the cross section for the collision of two initial-state hemisphere jets with masses less than some scale $m$. Using mass and hemisphere jets makes this infrared-safe cross section depend on only a single parameter, rather than say energy and angle cuts, like Steman-Weinberg jets. 

The  initial-state hemispheres are defined by the initial-state thrust axis. For a 2-body $e^+e^-$ initial state, the thrust axis is the same as the collision direction and both hemisphere masses are zero. For $e^+e^-\gamma$ initial states, the thrust axis aligns with the hardest of the 3 momenta. The two softer momenta are in one hemisphere and the hemisphere mass containing the single hard particle is zero. For simplicity, we ignore the region of phase space with $e^+e^-$ in the same hemisphere, as it is power suppressed and does not contribute an infrared divergence. Although we consider states with $m+2$ particles in the initial state, there are still at most 3 independent momenta, so we do not have to worry about the more complicated 4-body computation of the thrust axis and hemisphere masses. 

We calculate the cumulant total cross section, integrated over the phase space where both hemisphere masses are less than $\sqrt{\lambda} Q$. That is, $m_{\text{hemi 1}}^2 \le \lambda Q^2$ and $m_{\text{hemi 2}}^2 \le \lambda Q^2$. 
So at leading order
\be
\st_{00}(\lambda) = \sigma_0^d \delta(1-z)
\ee
where $\Gm = \left( \frac { 4 \pi e ^ { - \gamma_E } \mu ^ { 2 } } { Q ^ { 2 } } \right) ^ { \frac { 4 - d } { 2 } }$, 
$\sigma_0^d = \sigma_0 \frac{d-2}{2}\mu^{4 - d}$,
 $\sigma_0 =\frac{4 \pi g^2}{Q^2}$, and $z=\frac{m_Z^2}{Q^2}$ as before. 
The virtual correction $\st_{00}$ is the same as in Eq.~\eqref{eeZloop}.
\be
\st_{00}(\lambda)=\sigma_0^d  \frac{e^2}{\pi^2}\Gm \left\{-\frac{1}{4\eps^2} - \frac{3}{8 \eps}  + \frac{7\pi^2}{48} -1 \right\}\delta(1-z)
\ee
There is no $\lambda$ dependence as the virtual contribution always contributes. 

The only other contribution with no photons in the final state at this order has one photon in the initial state. The cross section is
\begin{multline}
\st_{10}(\lambda)=\sigma_0^d  \frac{e^2}{\pi^2}\Gm \left\{ 
\frac { 1 } {4 \eps^2} + \frac{3}{8 \eps} - \frac { 1 } { 4 } \ln^ 2  \lambda  - \frac { 3 } { 8 } \ln  \lambda - \frac { 5 \pi ^ { 2 } } { 48 } + \frac { 7 } { 8 }
+ \frac { 7 } { 32 } \ln ( 1 - 2 \lambda )
\right.
\\
\left.
+\lambda \left( \frac { 15 } { 16 } + \frac { 1 } { 2 } \ln  \lambda - \frac { 1 } { 2 } \ln ( 1 - 2 \lambda ) \right)  +\lambda^2 \left( \frac { 3 } { 16 } - \frac { 1 } { 8 } \ln  \lambda  + \frac { 1 } { 8 } \ln ( 1 - 2 \lambda ) \right)  - \frac { 1 } { 2 } \operatorname { Li } _ { 2 } ( 2 \lambda )
\right\} \delta(1-z)
\end{multline}
Note that the IR divergences of $\st_{00}$ and $\st_{10}$ exactly cancel, leaving $\ln \lambda$ and $\ln^2 \lambda$ terms, just as for final state jets. 

For $n>0$ photons in the final state, we have to be a little careful about the kinematics. 
For $Q \approx m_Z$, the jet masses can only be as large as roughly $m \lesssim  Q-m_Z$, thus $\lambda \lesssim (1-z)$. For values of $\lambda$ larger than this, the cumulant becomes $\lambda$ independent. The precise cutoff depends on the numbers of photons in the initial state and final state.  Explicitly $\st_{mn}$ becomes $\lambda$
independent for $\lambda > \frac{m}{2 n}(1-z)$.  For simplicity, we also take $z > \frac{1}{2}$ as we want the Born process to be $e^+e^- \to Z$ not $e^+e^- \to \gamma Z$. 

The various contributions for $\lambda > \frac{m}{2n}(1-z)$ are, for $m=n-1$,
 \begin{multline}
 \st_{n-1 , n } = \sigma_0^d \frac{e^2}{\pi^2} \Gm
 \left\{ \delta ( 1 - z ) \left( \frac { 1 } { 4\eps^2 } + \frac{\ln n}{\eps}- \frac { \pi ^ { 2 } } { 16 } + \frac { \ln ^ { 2 } n } { 2 }  \right) 
\right. \\
-  \frac { 1 } { \eps }   \frac { \left( 2 n ^ { 2 } - 2 n + 1 \right) z ^ { 2 } + 2 ( n - 1 ) z + 1 } { 4 n^2 } \left[ \frac { 1 } { 1 - z } \right] _ { + }
 \\
 \left.
+ \frac { \left( 2 n ^ { 2 } - 2 n + 1 \right) z ^ { 2 } + 2 ( n - 1 ) z + 1 } { 4 n ^ { 2 } } \left(  \ln \left( \frac{\left(n-1\right)z+1 }{n^3} \right) \left[ \frac { 1 } { 1 - z } \right] _ { + }
+2 \left[ \frac{\ln(1-z)}{1-z}\right]_+
\right)
 \right\}
 \end{multline}
 for $m=n$
 \begin{multline}
 \st_{n , n } = \sigma_0^d \frac{e^2}{\pi^2} \Gm 
 \left\{ \delta ( 1 - z )  \left( -\frac { 1 } { 2\eps^2 } -  \frac{2 \ln n}{\eps}+ \frac { \pi ^ { 2 } } { 8 } - \ln^2n \right) 
 +  \frac { 1 } { \eps } \frac { \left( 2 n ^ { 2 } + 1 \right) z ^ { 2 } - 2 z + 1 } {2 n^2 } \left[ \frac { 1 } { 1 - z } \right] _ { + }
 \right.\\
 \left.
 - \frac { 1 - z } { 2 n^2 } - \frac { \left( 2 n ^ { 2 } + 1 \right) z ^ { 2 } - 2 z + 1 } { 2 n ^ { 2 } } \left(  \ln \left( \frac{z }{n^2} \right) \left[ \frac { 1 } { 1 - z } \right] _ { + }
+2 \left[ \frac{\ln(1-z)}{1-z}\right]_+
\right) \right\}
 \end{multline}
 and for $m=n+1$,
 \begin{multline}
 \st_{n+1 , n } = \sigma_0^d \frac{e^2}{\pi^2} \Gm
 \left\{ \left( \frac { 1 } { 4\eps^2 } + \frac{\ln n}{\eps} - \frac { \pi ^ { 2 } } { 16 } + \frac { \ln^2 n } { 2 }  \right) \delta ( 1 - z )
 \right.
 \\
-\frac{1}{\eps}  \frac { \left( 2 n ^ { 2 } + 2 n + 1 \right) z ^ { 2 } - 2 ( n + 1 ) z + 1 } { 4 n^2 } \left[ \frac { 1 } { 1 - z } \right] _ { + }
 \\
 \left.
 + \frac { \left( 2 n ^ { 2 } + 2 n + 1 \right) z ^ { 2 } - 2 ( n + 1 ) z + 1 } { 4 n ^ { 2 } } \left(  \ln \left( \frac{\left(n+1\right)z-1 }{n^3} \right) \left[ \frac { 1 } { 1 - z } \right] _ { + }
+2 \left[ \frac{\ln(1-z)}{1-z}\right]_+
\right) \right\}
 \end{multline}
 For $\lambda \le \frac{m}{2n}(1-z)$, we find for $m=n-1$:
  \begin{multline}
 \st_{n-1 , n } = \sigma_0^d \frac{e^2}{\pi^2} \Gm
 \left\{- \frac{1}{\eps} \frac { \left( 2 n ^ { 2 } - 2 n + 1 \right) z ^ { 2 } + 2 ( n - 1 ) z + 1 } { 4 n^2 } \left( \frac { 1 } { 1 - z } \right)
 \right.
 \\
\left.
 + \frac {  (  n - 1 ) ( 1 - z ) - 2 n \lambda  } { 4 n ^ { 2 } ( n - 1 ) } + \frac { \left( 2 n ^ { 2 } - 2 n + 1 \right) z ^ { 2 } + 2 ( n - 1 ) z + 1 } { 4 n ^ { 2 } ( 1 - z ) } \ln \left( \frac { ( ( n - 1 ) z + 1 ) \lambda ( 1 - z ) ^ { 2 } } { n ^ { 2 } \left(( n - 1 ) ( 1 - z ) - n \lambda \right) } \right) \right\}
 \end{multline}
 for $m=n$
 \begin{multline}
 \st_{n , n } = \sigma_0^d \frac{e^2}{\pi^2} \Gm
 \left\{ \frac{1}{\eps} \frac { \left( 2 n ^ { 2 } + 1 \right) z ^ { 2 } - 2 z + 1 } { 2 n^2 } \left( \frac { 1 } { 1 - z } \right)
 \right.\\
 \left.
 + \frac { (1 - z ) - 4 \lambda } { 2 n ^ { 2 } } - \frac { \left( 2 n ^ { 2 } + 1 \right) z ^ { 2 } - 2 z + 1 } { 2 n ^ { 2 } ( 1 - z ) } \ln \left( \frac {  z \lambda ( 1 - z ) ^ { 2 } } { n ^ { 2 } ( 1 - z - \lambda ) }  \right) \right\}
 \end{multline}
 and for $m=n+1$,
 \begin{multline}
 \st_{n+1 , n } = \sigma_0 \frac{e^2}{\pi^2} \Gm
 \left\{ -\frac{1}{\eps} \frac { \left( 2 n ^ { 2 } + 2 n + 1 \right) z ^ { 2 } - 2 ( n + 1 ) z + 1 } {4 n^2 } \left( \frac { 1 } { 1 - z } \right)
 \right.
 \\
\left.
 + \frac {  ( n + 1 ) ( 1 - z ) - 2 n \lambda  } { 4 n ^ { 2 } ( n + 1 ) } + \frac { \left( 2 n ^ { 2 } + 2 n + 1 \right) z ^ { 2 } - 2 ( n + 1 ) z + 1 } { 4 n ^ { 2 } ( 1 - z ) } \ln \left( \frac { ( ( n + 1 ) z - 1 ) \lambda ( 1 - z ) ^ { 2 } } { n ^ { 2 } ( ( n + 1 ) ( 1 - z ) - n \lambda ) } \right) \right\}
 \end{multline}
 
 As with the total cross section, the IR divergences from these contributions cancel in triplets: $ \st_{n - 1 , n } + \st_{n , n} + \st_{ n+1 , n}$ is finite for any $n$. To see if the sum over $n$ converges, we look at the large $n$ behavior of the series. The asymptotic behavior for $n \gg 1$ is 
 \be
\st_{n - 1 , n } + \st_{n , n} + \st_{ n+1 , n} = \sigma_0^d \Gm \frac{e^2}{\pi^2}\times
\begin{cases}
-\frac{(1-z)^3}{6 z^2 n^4} + \cO(\frac{1}{n^6}), \qquad \lambda > 1-z \\[2mm]
-\frac{z(3\lambda^2 -4 \lambda(1-z) + 2 z^2-3z+1)}{2n^2(1-z-\lambda)^2} + \frac{3-6 \lambda}{2 n^2}+ \cO(\frac{1}{n^4}), \qquad \lambda < \frac { 1 - z } { 2 }
\end{cases}
 \ee
 Note that for $\lambda > 1-z$ the asymptotic behavior is the same as the total cross section, Eq.~\eqref{stot}, as expected, and that the sum converges for any $\lambda$. 
 
 While it is satisfying that the sum converges, we have be careful drawing too strong conclusions. As pointed out in~\cite{Lavelle:2005bt} for potential scattering, series like this one are not absolutely convergent. Summing in terms in a different order will give a different answer. For example, grouping by fixed number of initial state photons $\st_{m,m-1} + \st_{m,m} + \st_{m,m+1}$ the IR divergences still cancel in triplets, however there is a leftover uncancelled IR divergence $\st_{00}$. So the sum over $m$ is also IR divergent.
 Besides the ordering ambiguity, it is not at all clear that the cancellations and convergence will persist at higher order in perturbation theory or in QCD rather than QED. 
There is clearly much more to be understood, both computationally and physically, about initial state jets. 
 
\bibliography{IRfinite}

\bibliographystyle{utphys}

 \end{fmffile}

\end{document}